# Quantum Physics and Nanotechnology

## V.K. Nevolin

**Annotation**

Experimental studies of infinite (unrestricted at least in one direction) quantum particle motion using probe nanotechnologies have revealed the necessity of revising previous concepts of their motion. Particularly, quantum particles transfer quantum motion nonlocality energy beside classical kinetic energy, in other words, they are in two different kinds of motion simultaneously. The quantum component of the motion energy may be quite considerable under certain circumstances. Some new effects were predicted and proved experimentally in terms of this phenomenon. A new prototype refrigerating device was tested, its principle of operation being based on the effect of transferring the quantum component of the motion energy.

# Contents

# Introduction

At all the times new technologies have been conducive to the progress of science, nanotechnology being no exception. Nanotechnology is a new area of applied science dealing with fundamental properties of matter on the nanoscale and using them for the benefit of people. The mankind has the right to expect cardinal improvements in the quality of life thanks to the development and use of nanotechnology.

Experimental studies of infinite (unrestricted at least in one direction) quantum particle motion using probe nanotechnologies [1] have revealed the necessity of revising previous concepts of their motion. Particularly, quantum particles transfer quantum motion nonlocality energy beside classical kinetic energy, in other words, they are in two different kinds of motion simultaneously. The quantum component of the motion energy may be quite considerable under certain circumstances. Some new effects were predicted and proved experimentally in terms of this phenomenon.

A prototype refrigerating device where the cathode is cooled due to transfer of quantum energy component (Fermi energy) has been tested experimentally. Our calculations show that the efficiency of this device can be as high as 60%. We have also developed an experimental technique to determine the Fermi energy difference at electrodes. It is shown that the total energy of particles undergoing alpha decay differs from their kinetic energy by some percent. This result is important for developing precise alpha sources of heat and electricity.

A new physical effect revealing the possibility of quantum energy component enhancement is predicted. The matter is that the kinetic (thermal) energy of particles undergoing chemical and nuclear transformations can be decreased at the expense of quantum component enhancement. In this case we can speak about "cold reactions".

Some model problems of infinite particle motion have been solved; it helped to eliminate the existing theoretical problems in comprehension of some phenomena and strengthen our confidence that the new approach to description of infinite particle motion is more adequate. Understanding the applied significance of the suggested approach in describing infinite motion of quantum particles, the author popularized his ideas in some publications [1-3].

The author pays tribute to his teachers A.A. Kokin and V.M. Eleonsky for discussing basic approaches to the description of infinite quantum motion.

Additional literature offered by Yu.I. Bogdanov:

# 1. Background of the Problem

At the beginning of the 20th century some new experiments were accomplished, classical physics being incapable to explain their results. Actually, they gave birth to a new area of physics, quantum mechanics. A concept of wave function was introduced in quantum physics, which has no direct physical meaning but helps describe the time evolution of quantum systems. The square of wave function modulus represents the space-time probability density for the certain quantum system.

Quantum mechanics of infinite particle motion is certainly the most questionable area of the new physics. Every approach to the deduction of the quantum particle dynamics equation [1, 2], no matter how general it is, results in the Schrodinger equation. A classical formula for the kinetic energy $E$ of a free particle possessing a momentum $p$ and a mass $m$ is taken as the basis:

$$E = p^2/2m. \tag{1.1}$$

a concept of de Broglie wave is introduced

$$\Psi(p,t) = Ae^{i(\frac{\vec{p}\vec{r} - Et}{\hbar})}, \tag{1.2}$$

Thus we obtain the Schrödinger equation for a free particle which describes its space-time evolution in terms of the wave function $\Psi$:

$$i\hbar \frac{\partial \Psi}{\partial t} = \hat{H}\Psi \tag{1.3}$$

where the Hamiltonian for the free particle is of the form:

$$\hat{H} = (\hat{p})^2/2m = -\frac{\hbar^2}{2m}\Delta = -\frac{\hbar^2}{2m}(\frac{\partial^2}{\partial x^2} + \frac{\partial^2}{\partial y^2} + \frac{\partial^2}{\partial z^2})$$

$\hbar$ is Plank constant.

The Schrodinger equation is a complex one, having two corresponding real equations. As we mentioned before, the wave function is also a complex one having no direct physical meaning. The physical meaning can be assigned only to the probability density; it is this physical quantity that describes the space-time evolution of a paticle:

$$\rho(\vec{r},t) = \Psi \cdot \Psi^*, \tag{1.4}$$

$\Psi^*$ stands for the complex conjugate function.

At this stage we face the first contradiction. Introducing (1.2) into (1.4) we obtain the probability density to be a constant at any point of space, this fact being unaccountable. It means that the probability density for any free particle having momentum $\vec{P}$ is independent of coordinates and time, that is, constant in the whole space. That conclusion contradicts the existing experimental data. Any attempts to use a wave packet based on the superposition principle failed to eliminate the contradiction, as the packet spreads in all directions in the course of time. In connection with this fact one of the state-of-the-art methods to solve quantum problems of infinite motion is to describe the motion using the packet envelope on the time scale much less than the packet spreading time. Some other contradictions of infinite motion description based on de Broglie wave functions will be shown further.

The basic facts of infinite motion description in quantum mechanics, which are still unaccountable, originate from the contradiction mentioned. In our opinion, the reason for the situation is the refusal to describe quantum systems by means of physical quantities at the dawn of quantum mechanics. This is an expensive fee for the introduction of the unphysical wave function $\Psi$. The interpretation of quantum mechanics by means of physical variables, though, can help not only to eliminate the existing contradictions but also to predict new physical effects and verify them experimentally.

After the Schrodinger equation publication the same approach was proposed by E. Madelung. In 1926 he published quantum dynamics equations using physical variables, which were of quasi-hydrodynamic form. One of his equations turned out to be non-linear. Only in the 1950s did American physicist D. Bohm find the equations in the archives; later he made a considerable contribution to the hydrodynamic approach to the quantum system description [3,4]. Since then a non-linear method of quantum particlemotion description by means of physical variables having a physical meaning has been used to solve quantum problems. For example, it proved to be helpful in numerical calculations of quantum particle scattering [5]. Finally, the usage of quasi-hydrodynamic approach is reasonable if new experimentally verifiable results are or can be obtained.

Possibly one of the reasons why the quasi-hydrodynamic representation didn't take hold in quantum mechanics is that one of the equations is non-linear and difficult to solved analytically. However, there are not so many problems in quantum mechanics which can be solved analytically even with the linear Schrodinger equation.

The search of non-trivial solutions for infinite single-particle states led us to solving Schroedinger equation in hydrodynamic representation. Quantum hydrodynamic equations give us a possibility to describe infinite states of quantum particles sequentially. If needed, the results obtained can be verified by traditional Schrodinger equation solutions. The use of quantum hydrodynamic equations with physical variables enables us to look at the nature of infinite single-particle states in a different way.

## 2. Total Energy and Wave Function of Free Particle [2]

Unfortunately, in a number of textbooks on quantum mechanics formula (1.1) is considered as an expression for the total energy of a free particle. Let us rewrite it once more:

$$E = p^2/2m. \tag{2.1}$$

However, this expression discribes only the energy of translation motion and any quantum particle also takes part in a quantum motion, it is its inherent property, no matter what states it is in, finite or infinite. Thus any free particle is in two kinds of motion simultaneously revealing its wave-particle dualism with a certain amount of energy corresponding to each kind of motion.

Now we write down the expression for the Hamiltonian operator for a free particle of mass $m$:

$$\hat{H} = \hat{\mathbf{p}}^2/2m \tag{2.2}$$

In quantum mechanics it is accepted that only a quantum-mechanical average value of an operator corresponds to a real physical quantity. Thus, the energy of a particle can be written as

$$E = \langle \hat{H} \rangle = \langle \hat{\mathbf{p}}^2 \rangle/2m = \langle \mathbf{p} \rangle^2/2m + \langle (\delta\mathbf{p})^2 \rangle/2m \tag{2.3}$$

Here we assume

$$E = \langle \hat{H} \rangle = \int \Psi^* \hat{H} \Psi d\mathbf{r} \qquad \text{and} \qquad \langle (\hat{\mathbf{p}} - \langle \mathbf{p} \rangle)^2 \rangle = \langle (\delta\mathbf{p})^2 \rangle$$

From the above equations we can easily see that a quantum particle takes part in two kinds of motion: a translation motion, with the energy being equal to

$$E_k = \langle \mathbf{p} \rangle^2/2m$$

and a purely quantum motion, with the energy of quantum motion nonlocality being caused by momentum fluctuations

$$\delta\varepsilon = \langle(\delta\mathbf{p})^2\rangle/2m$$

Thus, the total energy of a particle equals

$$E = E_k + \delta\varepsilon. \qquad (2.4)$$

Let us now use the superposition principle of quantum states for a particle in two motions simultaneously and write down its wave function in the following form:

$$\Psi(\mathbf{r},t) = \frac{\sqrt{\rho_0}}{2}\left(e^{\frac{i(\mathbf{p}_1\mathbf{r}-E_1 t)}{\hbar}} + e^{\frac{i(\mathbf{p}_2\mathbf{r}-E_2 t)}{\hbar}}\right) \qquad (2.5)$$

We assume

$$\langle\mathbf{p}\rangle = (\mathbf{p}_1+\mathbf{p}_2)/2 \qquad \delta\mathbf{p} = (\mathbf{p}_1-\mathbf{p}_2)/2$$

$$E_1 = p_1^2/2m \qquad E_2 = p_2^2/2m \qquad E = (E_1+E_2)/2$$

From now on we designate $\langle\mathbf{p}\rangle = \mathbf{p}$. In the above mentioned terms the probability density of a particle in a finite motion will be given by equation

$$\rho(\mathbf{r},t) = \rho_0 \cos^2\left(\frac{\delta\mathbf{p}(\mathbf{r}-t\mathbf{p}/m)}{\hbar}\right) \qquad (2.6)$$

Here the initial phase of the wave is considered zero. In this case one of the peaks of the probability density function coincides with the classical position of the particle, the peak moving in the space with the momentum **p.** Using a larger number of wave functions to write a superposition describing a free particle motion results in a well-known problem of packet spatial time-spreading for every particle. With the total energy of a particle designated as $E$ and its average momentum designated as *p*, the wave function (2.5) can be rewritten in the form:

$$\Psi(\mathbf{r},t) = \sqrt{\rho_0}\cos\left(\frac{\delta\mathbf{p}(\mathbf{r}-t\mathbf{p}/m)}{\hbar}\right) e^{\frac{i(\mathbf{p}\mathbf{r}-Et)}{\hbar}} \qquad (2.7)$$

We can see from Eq. (2.7) that the plane wave amplitude is modulated by the harmonic function, its maximum propagating with classical velocity **p**/m. The spatial oscillation period obeys the following equations at any given time:

$$\delta p_x \cdot \delta x = 2\pi\hbar, \quad \delta p_y \cdot \delta y = 2\pi\hbar, \quad \delta p_z \cdot \delta z = 2\pi\hbar$$

It is easy to show that substitution of Eq. (2.7) for the wave function in the Schroedinger equation for a free particle leads us to the expression for the total energy in the form (2.3).

It will be shown further that the expression (2.6) for the free particle probability density is an analytical solution to quantum motion equations in quasi-hydrodynamic representation.

In the general case the probability density wave of a free particle (2.6) exhibits transverse-longitudinal oscillations, their wave vector being

$$\mathbf{k} = \delta \mathbf{p} / \hbar , \qquad (2.8)$$

with the frequency

$$\omega = (\delta \mathbf{p}/\hbar)(\mathbf{p}/m) = \mathbf{k}\mathbf{v}. \qquad (2.9)$$

It is essential that the dispersion law of such a particle is linear. Using expression (2.6) for the wave function we can qualitatively explain well-known experimental results for self-interference of a particle passing two slits [1]. It should be mentioned that to describe the infinite motion of a single particle the monograph [1] proposes a superposition of two wave functions after passing through the slits, to interpret the result of the interference.

Using (2.6) we can rewrite the energy conservation law for free particles in the form

$$E = E_k + (\hbar k)^2 / 2m \quad \text{or} \quad E = E_k + (\hbar \omega / 2)(\hbar \omega / 2 E_k) + \hbar^2 k_\perp^2 / 2m \qquad (2.10)$$

$k_\perp$ is the transversal (in relation to the motion direction) component of the particle wave vector. It can be seen that the quantum component of the particle free motion energy is of the wave nature and is connected, evidently, with the energy of probability density quantum oscillations. It should be noted that the frequency of probability density oscillations, according to (2.6) and (2.9) is doubled.

If we do not consider the transversal component of the momentum fluctuations ($k_\perp = 0$) and assume the quantum component of the particle energy to be equal to its kinetic energy $E_k = \hbar \omega / 2$ we shall obtain earlier postulates of quantum mechanics for particles with non-zero masses:

$$E = \hbar \omega \qquad \mathbf{P} = \hbar \mathbf{k}$$

These equations describe just a particular case of more general Eqs (2.10).

Reference:
1. The Physics of Quantum Information. Edited by D. Bouwmeister, A. Ekert, A. Zeilinger. Springer-Verlag Berlin Heidelberg 2000. P. 18.
2. V.K. Nevolin. On Motion Energy of Free Quantum Particles in Rarefied Beams. Ingenernaya Physika Journal. 2009. No. 5 P.20.

## 3. Quantum Mechanics Equations with Physical Variables

Let us now turn to quantum dynamics equations in quasi-hydrodynamic representation. As we mentioned before, they were probably first published by E. Madelung in 1926 after E. Schroedinger's equations and then by D. Bohm in the 1950s [1, 2].

We shall use the conventional Schrodinger equation for a particle with mass $m$ in an arbitrary potential field, without a spin or magnetic field:

$$i\hbar \frac{\partial \Psi}{\partial t} = \hat{H}\Psi, \quad \rho = \Psi\Psi^*. \tag{3.1}$$

The above written equation is complex and corresponds to a pair of equations in real space. One of these equations – the so called probability density conservation equation (or the continuity equation) – can be found in many books on quantum mechanics:

$$\frac{\partial \rho}{\partial t} + \frac{1}{m} div\, \mathbf{J} = 0, \tag{3.2}$$

the flux vector being equal to

$$\mathbf{J}/m = \frac{1}{2m}(\Psi^*\hat{\mathbf{P}}\Psi + \Psi\hat{\mathbf{P}}^*\Psi^*) \tag{3.3}$$

where $\hat{P}$ is the momentum operator. If we consider an infinite motion, not limited at least from one of the sides, there exists a macroscopic momentum $\mathbf{P}$

$$\hat{\mathbf{P}}\Psi = \mathbf{P}\Psi.$$

Thus, Eq. (3.1) can be rewritten in the form

$$m\frac{\partial \rho}{\partial t} + div\, \rho\mathbf{P} = 0. \tag{3.4}$$

Equation. (3.4) can be obtained by multiplying (3.1) and its complex conjugate equation by $\Psi$ and $\Psi^*$, respectively, and subtracting the products. The following dynamical equation is the result of summing the products (for more detailed derivation see App. 1):

$$\frac{\partial \mathbf{P}}{\partial t} = -\nabla(\frac{P^2}{2m} + U + \frac{\hbar^2(\nabla\rho)^2}{8m\rho^2} - \frac{\hbar^2\Delta\rho}{4m\rho}), \tag{3.5}$$

The set of Equations (3.4), (3.5) with the probability density $\rho(x, y, z, t)$ and momentum $\mathbf{P}$ is closed and equivalent to (3.1). It can be seen that (3.5) is quasi-hydrodynamic and non-linear, its form slightly differs from that in [2, 3].

If there is no macroscopic momentum for a particle, for example, in the area of tunneling, the set of Equations (3.4), (3.5) should be written with other variables. It will consist of Eq. (3.2) and equation

$$\frac{\partial (\mathbf{J}/\rho)}{\partial t} = -\nabla \left( \frac{J^2}{2m\rho^2} + U + \frac{\hbar^2 (\nabla \rho)^2}{8m\rho^2} - \frac{\hbar^2 \Delta \rho}{4m\rho} \right), \qquad (3.6)$$

If a quantum system of $N$ non-interacting particles each having its macroscopic momentum $P_n$ is considered the hydrodynamic equations are of the form:

$$\frac{\partial \mathbf{P}_i}{\partial t} = -\nabla_i \sum_{n=1}^{N} \left( \frac{P_n^2}{2m} + U_n + \frac{\hbar^2 (\nabla_n \rho)^2}{8m\rho^2} - \frac{\hbar^2 \Delta_n \rho}{4m\rho} \right), \qquad (3.7)$$

$$m \frac{\partial \rho}{\partial t} + \sum_{n=1}^{N} \nabla_n (\rho \mathbf{P}_n) = 0 \qquad (3.8)$$

It can be shown by means of Eqs. (3.7) and (3.8) that the probability density for a system of non-interacting particles is equal to the product of single-particle probability densities.

What kind of role does the quasi-hydrodynamic representation of quantum equations play? From our point of view, this representation checks the principle of wave function superposition, lays out the specifics of the superposition principle, prevents infinite summation of quantum states, unlike the wave packet. In case of infinite motion of quantum particles there exists a wave function corresponding to each component of the total energy.

The superposition (summation) of wave functions results not only in a new quantum state, but also in changing the total quantum system energy, as it can be seen from the Schroedinger equation.

Solving quantum equations in a quasi-hydrodynamic representation for infinite motion gives rise to a series of new experimentally verifiable physical effects.

**4. Infinite Motion of Quantum Particle in Quasi-hydrodynamic Representation**

Considering infinite motion of a particle with a macroscopic momentum $P$ in a stationary external field $U(r)$ the following set of equations should be solved:

$$\frac{\partial \mathbf{P}}{\partial t} = -\nabla\left(\frac{P^2}{2m} + U + \frac{\hbar^2(\nabla\rho)^2}{8m\rho^2} - \frac{\hbar^2\Delta\rho}{4m\rho}\right), \tag{4.1}$$

$$m\frac{\partial\rho}{\partial t} + \operatorname{div}\rho\mathbf{P} = 0, \tag{4.2}$$

Taking into account that $\mathbf{P}=\mathbf{P}(r)$ in a stationary problem, one can see from (4.1) that the total energy $E$ of the particle is conserved:

$$E = \frac{P^2}{2m} + U + \frac{\hbar^2(\nabla\rho)^2}{8m\rho^2} - \frac{\hbar^2\Delta\rho}{4m\rho} = const \tag{4.3}$$

The quantity $\delta\varepsilon(r) = \frac{\hbar^2(\nabla\rho)^2}{8m\rho^2} - \frac{\hbar^2\Delta\rho}{4m\rho}$ will be further referred to as the energy of quantum motion fluctuations or the quantum component of the total energy. Thus, we can write

$$E = \frac{P^2}{2m} + U + \delta\varepsilon(\mathbf{r}). \tag{4.4}$$

The additive quantity $\delta\varepsilon(\mathbf{r})$ can be small but it is essentially non-zero for quantum particles, otherwise, the particle loses the quantum essence of its motion. On the other hand, the existence of $\delta\varepsilon(r)$ changes the space-time distribution of the probability density.

Therefore, a freely moving particle possesses, according to (4.4), not only the kinetic energy but also the energy of quantum movement fluctuations which is variable across the space.

Now let us find an analytical solution to the set of Equations (4.2), (4.3) for a free particle, when ($U(r)=0$ and $\mathbf{P}=const.$). The solution $\rho = const$ is trivial and leads to the conservation law for a classical particle, thus, we shall abandon it.

The general solution to Eq. (4.2) can be written in the following form:

$$\rho = \rho(\mathbf{r} - t\mathbf{P}/m). \tag{4.5}$$

Solving Eq. (4.3) is rather complicated, or, at least, one can verify the result written down (see App. 3):

$$\rho(r,t) = \rho_0 \cos^2(\delta p(r - tp/m)/\hbar), \qquad E = p^2/2m + \delta\varepsilon, \qquad \delta\varepsilon = (\delta p)^2/2m. \tag{4.6}$$

We have already discussed the main features of the solution (4.6) in section 2; now we should mention another important result. If the vectors $\delta\mathbf{p}$ and $\mathbf{p}$ are collinear a quantum particle can be found in so called "needle states", its transversal position is defined strictly. In the general case the value and direction of the vector $\delta\mathbf{p}$ for a free particle depend on its origin. For example, if an electron is tunneling from the top of the Fermi surface normally to the autocathode the transversal components of $\delta\mathbf{p}$ equal zero.

## 5. Thermal Effect of Autoelectronic Emission on Anode [4]

Now we shall cite experimental evidence in favour of the quantum energy component existence. During field emission of electrons from the cathode a certain amount of heat should be generated in the anode which was brought by accelerated electrons determined by the emission current strength and the voltage applied between the electrodes. But actually the amount of heat generated in the anode also depends on the difference of Fermi energies of the anode and cathode; this process was not observed before due to the peculiarities of previous experiments [1].

The essence of the effect is as follows. During the tunneling process through the triangular barrier in the external electric field a Fermi electron leaving the cathode carries away some energy of quantum motion fluctuations equal to the Fermi energy (we consider metallic electrodes). Then an electron moving in an accelerating field gains its energy of translation motion. Having penetrated into the anode an electron loses its energy as it is converted into heat until the electron reaches the Fermi surface of the anode. If the Fermi energies of the cathode and anode are different the amount of heat generated will differ from the expected.

Let us now formulate the criteria of observing the effect. To indicate the field emission process (through the triangular barrier) the voltage $U$ applied must exceed the biggest work functions of the cathode and the anode: $eU > \max(e\varphi_1, e\varphi_2)$. However, the voltage $U$ should not much exceed the Fermi energies of electrons of electrodes ($eU \leq \varepsilon_{f1}, \varepsilon_{f2}$), otherwise the thermal effect becomes vanishingly small. In previous experiments high voltages were commonly applied [1]. It is necessary to provide strong electric fields (for the field emission process to start the field strength should reach the value of $10^6$–$10^7$ V/cm). At $U=10$ V the interelectrode distance must not exceed 10 nm to provide the required field strength. All the necessary conditions can be implemented in scanning tunnel microscopy [2].

Let us estimate the value of the supposed effect. The energy of an electron leaving the cathode equals

$$E = p_1^2/2m + \varepsilon_{1f}, \tag{5.1}$$

where $\varepsilon_{1f}$ is the Fermi energy of the cathode. The energy of electron which has reached the anode is

$$E = p_2^2/2m - eU + \varepsilon_{2f} \tag{5.2}$$

where $\varepsilon_{2f}$ is the Fermi energy of the anode.

The kinetic energy of the electron in the anode which will be transformed into heat equals

$$(p_2)^2/2m = eU + (p_1)^2/2m + \varepsilon_{1f} - \varepsilon_{2f} \tag{5.3}$$

We shall neglect the thermal tailing of the electron energies in the cathode, in comparison with its Fermi energy, thus we consider the initial electron momenta infinitesimal $p_1 \succ 0$. Hence, Eq. (5.3) takes the form

$$(p_2)^2 / 2m = eU + \varepsilon_{f1} - \varepsilon_{f2}.$$

The relative heat generation in the anode as a function of the voltage applied is thus described by the following equation:

$$\Delta Q / Q = 1 + (\varepsilon_{f1} - \varepsilon_{f2}) / eU, \qquad (5.4)$$

where $Q = IU$ is the "classical" amount of heat generated. It can be seen from Eq. (5.4) that an effect of over- or underheating of the anode can be observed, depends on the difference of Fermi energies of the electrodes. It is only in a particular case of identical electrodes that the heat generation process is classical. Our next goal is to prove that quasi-classical electrons moving in external field after tunneling transfer the energy of quantum motion fluctuations; being equal to the cathode Fermi energy in this particular case.

The idea of the experiment is as follows. The substrate in a single-point tunneling device represents a plane microthermocouple. The probe of the scanning tunnel microscope is brought close to the thermocouple junction and the substrate temperature variation is measured at the given values of voltage and autoelectronic current applied to the substrate. As the temperature distribution from the point heat source in the near-surface area of the substrate is proportional to the voltage applied and the current strength, the graph $\Delta T(IU)/IU$ versus the voltage $U$ is universal for this case and clarifies the situation. If this dependence remains constant there is no effect (the heat generation is classical), otherwise we expect qualitative agreement with (5.4).

In the experiment electrochemically sharpened tungsten probes made of wire of diameter $d=1$ mm were used; the probe tip radius was about 20 nm. Tungsten work function, according to the reference data, equals $\varphi_1 = 4.5$ eV, the Fermi energy was expected not less than $\varepsilon_{f1}=14.5$ eV. The interelectrode voltage didn not exceed 8 V, which was lower than Fermi energies of the electrodes. A chromel-alumel thermocouple made of wire of diameter 190 $\mu$ m was used as an anode, the thermocouple was T-shape scarf-welded. The substrate itself was a flat alumel wire ground and polished down to 20 $\mu$ m thick. It was placed over the edge of the chromel wire. The expected value of the Fermi energy of the alumel substrate (95% Ni, the residue: Al, Si, Mn) was $\varepsilon_{f2}=11.7$ eV, the expected work function $\varphi_2 = 4.5$ eV. The difference between Fermi energies of the electrodes is such that the anode should be relatively overheated.

In the course of experiment some problems arose, including considerable fluctuations of autoemission current known since R. Young's topographiner [3] and time-drift of thermocouple EMF as the measurements were conducted close to its response limit.

It required quick measurements only in a few points in each experiment. The maximum value of thermocouple EMF reached 4 $\mu$V, which corresponds to the junction heating up to 0.1 K in accordance with the calibration scale. At the same time the substrate surface under the electron beam was heated by dozens of degrees. To limit and to measure the current a resistor of 100 kOhm was introduced into the circuit. The current strength reached the value of 10 $\mu$A at the voltage of 7.8 V, which could result in resistive heating of the probe tip, proportional to the square of the current flowing. Due to the resistive heating some additional thermionic current between the electrodes is possible, which decreases the effect as the thermions transfer mainly the transversal component of Fermi energy. The potential amount of heat transferred radiatively to the substrate, caused by heating the tip of a small area is many orders of magnitude less than the heat generation caused by the difference of Fermi energies at the electrodes at the given value of current. When the tip is heated over $T$=373<K the adsorbate (consisting mainly of water molecules) falls off and with the interelectrode distance being much less than the air molecule free path the molecular heat transfer doesn not exceed 10% of the expected effect in the worst case (the value of current strength equals 50 $\mu$A).

Fig. 1 shows experimental points of dependence of the ratio of EMF variation to the power generated at the anode $\Delta E / JU$ from reciprocal value of the voltage applied $1/U$. The points are plotted for different probes on different dates and for different points on the substrate. Within the range of uncertainties a universal dependency is obtained that can be approximated with a straight line having an evidently negative slope, which corresponds to the expected additional overheating of the anode. The massive thermocouple junctions prevented obtaining a steeper slope of this dependence.

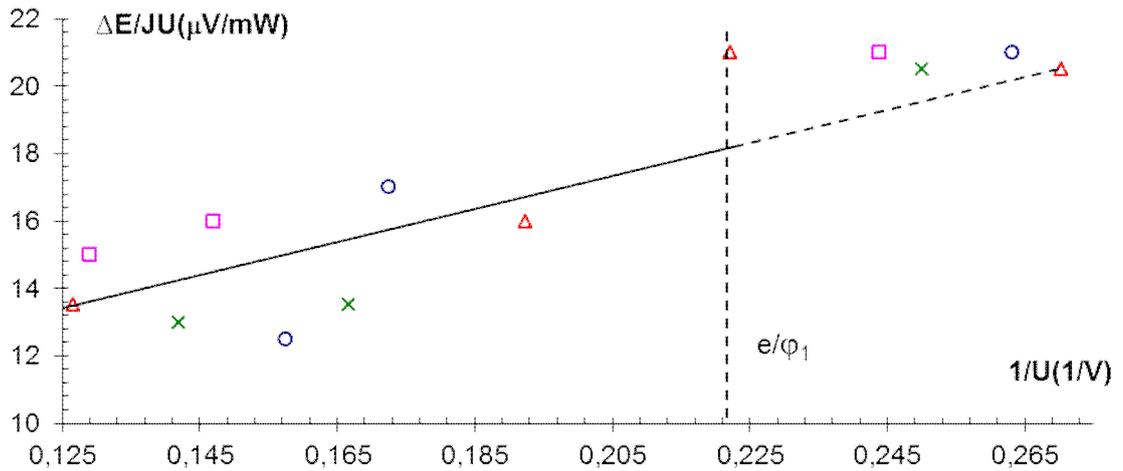

Figure 1

Hence the thermal effect is proved experimentally and its consequences can be analyzed. One of the possibilities is considered in the Appendix.

Reference

1. L.N. Dobretsov, M.V. Gomoyunova. Emission Electronics. M. Nauka 1966.402p.
2. V.K. Nevolin. Probe Nanotechnologies in Electronics. M. Tekhnosila. 2005. 148 p.
3. Young R., Ward J., Your R. Phys. Rev. Lett. 1971.V. 27, N14  P.922-924; Rev. Sc. Instr. 1972,. V. 43. N7.  P. 999-1011.
4. V.K. Nevolin. Thermal Effect on Anode at Field Emission. Technical Physics Letters, 2006, V.32, No. 12, pp. 1030-1032.

## 6. Effect of Anode Cooling at Field Emission [5]

At field emission of electrons from the cathode a certain amount of Joule heat should be generated in the anode which is brought by accelerated electrons in accordance with the emission current strength and the voltage between the electrodes. It seems reasonable that anode must always be heated. But actually, if we take into consideration the quantum component of the electrons energy and the ratio of Fermi energies of the electrodes to the voltage applied, we shall come up to the idea that the anode can be cooled. The effect was not observed before as high interelectrode voltages (much higher than Fermi energies) were used in the experiments.

The aim of the following section is to prove experimentally the possibility of anode cooling at field emission from the cathode.

The essence of the effect under consideration is elucidated by Fig. 1.

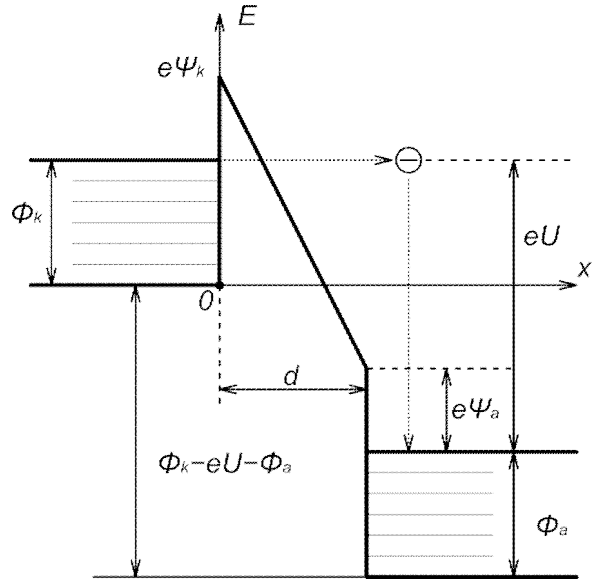

Figure 1. Band diagram of an electron tunneling from the cathode (left) to the anode (right) $e\Psi_k, e\Psi_a$ are electron work functions of the cathode and anode, respectively, $\Phi_k, \Phi_a$ are Fermi energies of the cathode and anode, respectively, $eU$ is the energy, gained by electrons in the external field $U$, $e$ is the electron charge, $d$ is the interelectrode distance

While tunneling through the triangular barrier in the external electric field a Fermi electron leaving the cathode carries away a quantum component of motion energy equal to Fermi energy (we consider metallic electrodes). Moving then in an accelerating field the electron gains kinetic energy. When the electron penetrates into the anode its total energy changes until the electron reaches the Fermi surface of the anode. If the Fermi energy of the anode differs from that of the cathode, the amount of heat generated in the anode will be different from the expected (given by Joule's law [1]).

To estimate the effect quantitatively one should have an expression for the energy of an electron moving in the interelectrode space. In quasi-hydrodynamic representation [2, 3] dynamical equations for infinite motion of a particle with mass $m$ in arbitrary external field $W(\mathbf{r}, t)$ are written in the form:

$$m \frac{\partial \rho}{\partial t} + \text{div}\, \rho\, \mathbf{p} = 0 \qquad (6.1)$$

$$\frac{\partial}{\partial t} \mathbf{p} = -\nabla \left( \frac{p^2}{2m} + W + \frac{\hbar^2 (\nabla \rho)^2}{8m\rho^2} - \frac{\hbar^2 \Delta \rho}{4m\rho} \right) \qquad (6.2)$$

where $\rho(\mathbf{r},t)$ is spatial-time distribution of particle probability density, $p(\mathbf{r},t)$ is the macroscopic momentum of the particle. The external field being stationary the total energy of the particle $E$ is conserved and hence, using Eq. (6.2), we can write down an analog of the Bernoulli invariant:

$$E = p(\mathbf{r})^2/2m + W(\mathbf{r}) + \delta\varepsilon(\mathbf{r}) = const, \qquad (6.3)$$

where $\delta\varepsilon = \dfrac{\hbar^2(\nabla\rho)^2}{8m\rho^2} - \dfrac{\hbar^2\Delta\rho}{4m\rho}$ is a quantum additive component of the total energy of the particle.

As the thermal component of the electron energy is much less than the Fermi energies of the electrodes we shall neglect it further. Thus the quantum component near the cathode is $\delta\varepsilon(0) = \Phi_k = E$ (see Fig. 1). The energy $E$ of the electron in the external electric field with potential $U$ which has reached the anode and has Fermi energy of the anode equals

$$E = p_a^2/2m - eU + \Phi_a \tag{6.4}$$

The relative heat generation in the anode as a function of the voltage applied is thus described by the following equation:

$$\Delta Q/Q = 1 + (\Phi_k - \Phi_a)/eU \tag{6.5}$$

where $Q=IU$ is the amount of Joule heat generated at the anode. From Eq. (6.5) one can see that in a certain range of voltages the amount of heat generated can be negative $\Delta Q < 0$ and the anode will be cooled despite the Joule heat, if the Fermi energy of the anode exceeds that of the cathode $\Phi_k - \Phi_a < 0$.

Let us now formulate the criteria of observing the effect. To indicate the field emission process (through the triangular barrier) the voltage $U$ applied must exceed the biggest work functions of the cathode and the anode: $eU > \max(e\Psi_k, e\Psi_a)$. In this case every electron as such moves in an accelerating electric field in a certain area of the interelectrode space and transfers the quantum component of the energy according to Eq. (3). On the other hand the voltage applied must not exceed the difference of Fermi energies of the electrodes ($eU < \Phi_a - \Phi_k$) to provide cooling. Hence, the range of applied voltages where the anode cooling effect can be observed can be presented as:

$$\max(e\Psi_k, e\Psi_a) < eU < \Phi_a - \Phi_k$$

To obtain appreciable autoelectronic current the electric field strength near the cathode should be about $10^7$ V/cm. At the voltage of some volts the interelectrode distance must be about 1nm. All the required conditions can be fulfilled in scanning tunnel microscopy [4].

The idea of the experiment is as follows. There was used scanning tunnel microscope Solver P47, whose tunnel head was upgraded to provide a range of set currents up to 50 $\mu A$ maintained by the feedback.

A plane thermocouple was used as a substrate. The tunnel probe was brought close to the thermojunction and the thermocouple EMF was measured at the varying substrate temperature and set values of the applied voltage and the substrate autoelectronic current. The temperature increase as a function of radius $r$ from the axis of electron beam to a certain point of the substrate is proportional to the flowing current and the voltage applied [4]:

$$\Delta T(r) = \frac{UI}{4\pi kl}\left(2\frac{l}{r}(1-e^{-r/l}) - e^{-r/l}\right) \qquad (6.7)$$

where $U$ is the voltage applied to electrodes, $I$ is the tunneling current, $k$ is the thermal conductivity coefficient of the substrate, $l$ is the free path of electron inelastic scattering in the substrate. According to Eq. (7) the ratio $\Delta T/Q$ remains constant during the Joule heating within small temperature variations, when the substrate material coefficients can be considered constant. But if we take into account the quantum component of t energy in accordance with (5) the variation of voltage $U$ can result in: anode cooling in case the inequality (6) is fulfilled; constant anode temperature at some voltage $U0$ when $eU_0 = \Phi_a - \Phi_k$; and finally, anode heating if $U>U_0$;

In the experiment mechanically sharpened silver, copper and gold alloy probes were used. The Fermi energy calculation was based on the valence electron concentration.

A chromel-alumel thermocouple made of wire of diameter 190 $\mu$ m was used as an anode, the thermocouple was T-shape butt-welded. The substrate itself was a flat alumel wire. The junction of two wires was made flat by grinding and polishing, with minimal contact area. The value of Fermi energy expected for the alumel substrate was $\Phi_a$ =11.7 eV (alumel alloy contains 95% of Ni, the rest is Al, Si, Mn); the expected value of the electron work function was $e\Psi_a$ =4.5 eV for Ni.

During the experiments some difficulties were encountered, and namely, considerable fluctuations of autoelectronic current known before, fluctuations and time-drift of thermocouple EMF as the measurements were conducted close to its response limit [1].

The fact that a strong electric field causes mutual attraction electrodes and their plastic yielding added some more problems. The value of the electric field strength causing plastic deformation of electrodes can be estimated using Eq. [4]

$$E_0 = 2.1 \cdot 10^3 \cdot \tau^{1/2} \text{ V/cm,} \qquad (6.8)$$

where $\tau$ is the strain resulting in plastic deformation.

According to the reference data $E_0$ =0.94 - 1.15*$10^7$ V/cm for the silver probe and $E_0$ =1.9* $10^7$ V/cm for the alumel (nickel) substrate. At strong fields sufficient for appreciable field emission to occur, plastic yielding of the electrodes (especially of the probe) was observed resulting in short-circuiting of electrodes with imprints in the form of hills left on the substrate. These hills were observed experimentally by scanning the substrate in the tunneling mode.

To process the experimental data in accordance with Eqs. (6.5) and (6.7) the following formula was used:

$$\Theta = \Theta_0 + \alpha I (U_a - \Delta\Phi/e),$$

where $\Delta\Phi = \Phi_a - \Phi_k$, $\Theta$ is the thermocouple EMF caused by autoelectronic current $I$ flowing between the probe and the substrate, $\Theta_0$ - is the initial thermocouple EMF value, $\alpha$ is the thermocouple sensitivity factor, not less than 0.07 $\mu V/\mu W$ in our case. The value of $\alpha$ depends on the probe position relative to the junction. The interlectrode voltage $U_a$ is the sum of the voltage across the spacing $U$ and the voltage drop on the electrodes of the total resistance $R$:

$$U_a = U + IR$$

The second summand results in generating additional amount of heat in the substrate decreasing the effect of its cooling. To control the resistive heating one had to change the current polarity, the voltage remaining the same. The substrate played the role of cathode and the cooling effect vanished; the substrate temperature was controlled by means of thermocouple readings. At small currents (about some microamperes) the contribution of this summand to the total heat generation was negligible. The experimental results are presented in the table below.

Table: Experimental results

| Electrode | $\Phi$, eV | $e\Psi$, eV | $\Delta\Phi$, eV calc. | $\Delta\Phi$, eV, experim. |
|---|---|---|---|---|
| Ag | 5.5 | 4.3 | 6.2 | $5.6 \pm 1.4$ |
| Au alloy, 58.5% | >5.5 | <5.1 | <6.2 | $4.2 \pm 1.1$ |
| Cu | 7.0 | 4.4 | 4.7 | $5.3 \pm 1.3$ |
| Alumel Ni, 95% | 11.7 | 4.5 | substrate | substrate |

One can see that experimental differences of Fermi energies $\Delta\Phi$ for silver and copper probes and an alumel substrate are in a good agreement with the calculated values within the uncertainty range. The probe fabricated from jewelry gold alloy (60% of Au, the rest is Cu) showed a lower value of $\Delta\Phi$ as in this case copper makes a sufficient contribution to the Fermi energy.

Thus, we have demonstrated the effect of anode cooling during autoelectronic emission; the main difference from the Pelletier effect is the requirement of the ineterelectrode spacing which enables electrons to tunnel from the cathode with the Fermi energy, then to gain energy in an external electric field necessary to transfer chargesnd to deliver the total energy to the anode.

The experiment confirms the additivity of the quantum energy component for particles in infinite motion in accordance with Eq. (6.3).

The new concept of quantum particles motion allows one to develop tunnel refrigerating devices with unprecedented theoretical cold outcome up to 60% of the power consumed. The existence of the interelectrode spacing, especially if the air is pumped out from there makes it possible to decrease useless heat outflow, unlike known thermal electric converters, and not to lower the device efficiency too much. [6]. Refrigerating elements of this kind can be realized only with nanotechnologies. Fig.2 shows a sketch of such element. The cathode (1) represents a conducting electrode with multi-walled nickel-headed carbon nanotubes placed perpendicularly to the electrode surface. The nickel heads usually appear on nanotube ends in the growth process; one should only unseal them. The anode (3) is covered with a conductive graphite film and separated from the cathode (1) with an electrothermal-insulating spacer (2). The diameter of the nanotubes must not exceed 30 nm, the distance between them must be at least half of their height. The interelectrode distance can reach many microns. The more the number of nanotubes on a unit area of the cathode the more cold is generated per the same unit area. It should be noted that refrigerating devices based on nanotechnologies which save electric energy are in demand in every house.

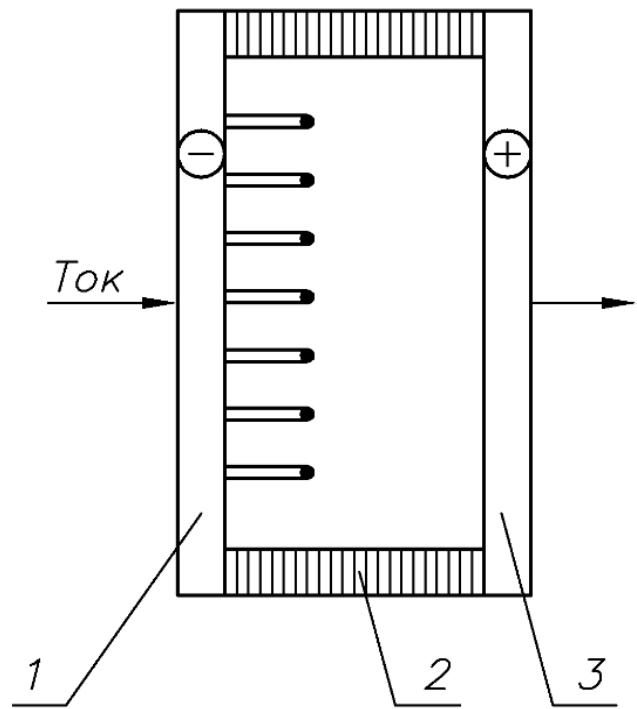

Figure 2.- Sketch of the autoemission refrigerating device: 1 – cathode with multi-walled nickel-headed carbon nanotubes; 2 - insulator; 3- anode with carbon conductiive coating.

Reference:
1. V.K. Nevolin. Thermal Effect on Anode at Field Emission. JPL Letters. Technical Physics Letters, 2006, V.32, No. 12, pp. 1030-1032.
2. S.K., Ghosh, B.M Deb. Densities, Density-Functionals and Electron Fluids. Physics Reports (Review Section of Physics Letters). 1982. V. 92, N1. P. 1-44.

### 7. Heat Emission by Alpha-Sources [4]

The fact that the emission of thermal energy at alpha-decay is somewhat higher than it can be expected based on kinetic energy experimentally measured with mass-spectrometers (to a high accuracy, as a rule) should be taken into consideration while designing and exploiting precision heat alpha-sources.

Quantum particles which are in infinite motion feature a fundamental property of transferring quantum motion nonlocality energy beside their kinetic energy. This additive energy is closely associated with the origin of particles and may be quite appreciable in experiments.

Naturally, the existence of quantum nonlocality energy with particles in infinite motion must show itself in other phenomena as well, for example, at charged particles tunneling from nuclei. In particular, charged particles at alpha-decay must carry away, beside kinetic energy, the energy of quantum motion nonlocality which can be measured as a difference between the total energy of particles, thermalized in the environment with given properties, and the kinetic energy of incoming particles. Presence of the charge makes it possible to measure the current and the kinetic energy of particles in the transverse magnetic field. Thus, if we know the kinetic energy of outgoing particles for alpha-sources and have calculated heat emission based on their kinetic energy in the environment with given properties, we will notice, that the amount of heat emitted will be somewhat larger due to additional thermalization of quantum motion nonlocality energy.

Our objective is to evaluate the amount of quantum motion nonlocality energy carried away by alpha-particles from nuclei.

In theory, presence of additive quantum nonlocality energy can be proved if we write an equation of quantum particle motion by means of physically meaningful quantities. This is so-called quasihydrodynamic representation, much contributed to by D. Bohm in the 50s of the last century (see review [1]).

For the purpose of certainty we will bear in mind that alpha particles move in Coulomb field of a nucleus $U = 2Ze^2/r$, where $Ze$ is the charge of the daughter nucleus. The equations of motion in quasihydrodynamic representation will be written as follows [1, 2]:

$$\frac{\partial \rho}{\partial t} + div\mathbf{J} = 0 \qquad (7.1)$$

$$m\frac{\partial}{\partial t}(\mathbf{J}/\rho) = -\nabla\left(\frac{mJ^2}{2\rho^2} + U + \frac{\hbar^2(\nabla\rho)^2}{8m\rho^2} - \frac{\hbar^2\Delta\rho}{4m\rho}\right), \qquad (7.2)$$

where $m$ is the reduced mass of a particle. The system of equations (7.1), (7.2) with probability density $\rho$ (**r**, t) and particle probability flow density $\mathbf{J}(r,t)/m$ is closed. If the notion of macroscopic momentum **p** can be introduced while describing quantum particle motion, then $\mathbf{J}/\rho = \mathbf{p}/m$.

When a particle is in infinite motion in a stationary external field with a macroscopic momentum **p** its total energy E is constant, then we have a Bernoulli invariant analog from (7.2):

$$E = \frac{p^2}{2m} + U + \delta\varepsilon(\mathbf{r}), \qquad (7.3)$$

where $\delta\varepsilon = \frac{\hbar^2(\nabla\rho)^2}{8m\rho^2} - \frac{\hbar^2\Delta\rho}{4m\rho}$. This value can be called the energy of quantum motion nonlocality ("quantum-mechanical" potential according to D. Bohm). The presence of $\delta\varepsilon$ can change radically the spatial-temporal distribution of probability density. In particular, a particle can be found in so-called "needle" states when it is localized strictly transversely. We can show it looking at a particle moving far from the centre of force, when $U \to 0$

A nontrivial solution to (7.1), (7.3) можно can be written as follows:

$$\rho(\mathbf{r},t) = \rho_0 \cos^2(\delta\mathbf{p}(\mathbf{r} - t\mathbf{p}/m)/\hbar), \qquad E = p^2/2m + \delta\varepsilon, \qquad \delta\varepsilon = (\delta p)^2/2m \qquad (7.4)$$

The motion of a free particle can be described using the previous language as a stable superposition of two plane wave functions which should be postulated:

$$\Psi(\mathbf{r}, t) = \frac{1}{2}\sqrt{\rho_0}\,(exp\ i(\mathbf{p_1 r} - E_1 t)/\hbar + exp\ i(\mathbf{p_2 r} - E_2 t)/\hbar) \qquad (7.4a)$$

where the free particle momentum **p** and its energy $E$ are defined by equations:

$(\mathbf{p_1} + \mathbf{p_2})/2 = \mathbf{p}, \qquad (E_1 + E_2)/2 = E, \qquad \delta\mathbf{p} = (\mathbf{p_1} - \mathbf{p_2})/2$

The simple relationship between the quantum nonlocality energy of a free particle $\delta\varepsilon_\infty$ and the mean value of its kinetic energy $E_k = p^2/2m$ can be found from (7.4) in case of needle states, if the typical problem time $\tau_0$, for example, the oscillation period of probability density in time, is introduced.

$$E = E_k + \delta\varepsilon_\infty, \delta\varepsilon_\infty = \frac{\pi^2 \hbar^2}{4 E_k \tau_0^2} \tag{7.5}$$

If the kinetic energy of a particle in (7.5) is small, the quantum motion nonlocality energy may exceed its kinetic energy, which is observed at field emission [5].

When a charged particle tunnels from a nucleus and moves in the field of the force center only a part of its total energy converts into kinetic energy. Let us consider a particle tunneling through the Coulomb barrier in the model of a rectangular potential pit with energy $E$ [3]. We shall try to solve a dynamic problem of particle tunneling in time in quasihydrodynamic representation [4].

In the field of tunneling we shall solve the system of equations (7.1), (7.2), using an invariant:

$$E = \frac{mJ^2}{2\rho^2} + U + \frac{\hbar^2 (\nabla \rho)^2}{8m\rho^2} - \frac{\hbar^2 \Delta \rho}{4m\rho} \tag{7.6}$$

We suppose the energy of tunneling particles to remain constant in stationary potential fields, then $\partial/\partial t\, (J/\rho) = 0$, and $J/\rho$ depends on the coordinate. As it can be seen from (7.6) a tunneling particle moves in some self-consistent potential field. Further we shall simplify the problem tending essentially to quasiclassical approximation. Considering only one-dimensional motion in the field of plane Coulomb barrier and ignoring spatial curvature of the barrier (the approximation degree will be evaluated below), we can find the solution to equation (7.1):

$$J = J\left(\frac{t - \int \Phi(r)\,dr}{\tau_0}\right),\ \Phi(r) = \rho(r,t)/J(r,t)$$

Then the solution to (7.6) can be written as follows:

$$\rho(r,t) = \Phi(r) J_0 \exp\left(\frac{t - \int \Phi(r)\,dr}{\tau_0}\right)$$

The sign choice in front of the exponent will be clear from the further explanation. We obtain the following expression:

$$\varphi^4 + 2\varphi\varphi'' - (\varphi')^2 - 4\varphi^2 \varphi' + \beta(r)\varphi^2 - \alpha^2 = 0, \tag{7.7}$$

where $\varphi = \Phi/\tau_0, \beta(r) = (E - U(r))8m/\hbar^2, \alpha^2 = 4m^2\tau_0^2/\hbar^2$ the prime in $\varphi$ denotes differentiation with respect to the coordinate. Considering coordinate derivatives in (7.7) to be small we have the following in zero approximation:

$$\varphi(r) = \left((\beta^2/4 + \alpha^2)^{1/2} - \beta/2\right)^{1/2} \tag{7.8}$$

It should be noted that (7.8) is the exact solution for the case of a plane rectangular barrier of finite width. The solution in the general case approximates the arbitrary-form barrier with a set of plane rectangular barriers, infinitely small in width.

The barrier permeability is defined as the ratio of probability flow density in the extremum point of self-consistent potential at some time to the initial probability flow. The permeability here must satisfy the extreme cases. If the barrier width tends to zero its permeability tends to one, and the time when a particle as it is can be located on top the barrier must be counted off from zero. Naturally, the permeability of the barrier must tend to zero with increasing its width. Then the permeability of a plane potential pit with a Coulomb potential barrier can be written as quadratures:

$$D = \exp(-\int_{r_0}^{r_1} \varphi dr - \int_{r_1}^{r_2} \varphi dr) \qquad (7.9)$$

Here $U(r_0) = U(r_1) = E$. The Coulomb barrier can be thought of as approximately plane if the de Broglie wavelength $\lambda$ of tunneling particles is considerably less than the Coulomb barrier radius $r_1$

$$\lambda / r_1 = \pi \hbar (E)^{1/2} / Ze^2 (2m)^{1/2} \ll 1$$

This circumstance was used earlier in [3] as well as in our solution.

The self-consistent tunneling process in accordance with (7.9) can be provisionally presented as consisting of two parts: overcoming the area $r_1 - r_0$ wide, when $U(r) \geq E$, and overcoming the area $r_2 - r_1$ ($E > U(r)$), when the particle is formed as it is with some energy of quantum motion nonlocality and can be in infinite motion after passing the point $r_2$, see Fig. 1.

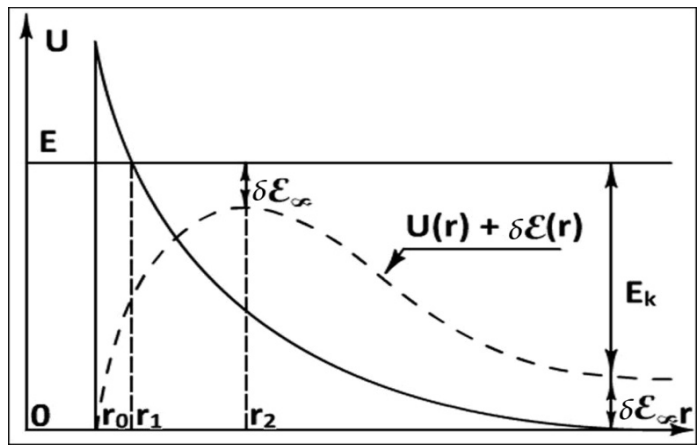

Fig. 1. Schematic diagram of a charged particle tunneling from a nucleus.

The point $r_2$ corresponds to the minimum probability density flow energy.

$$E_j = mJ^2(r_2)/2\rho^2(r_2) = \min = \delta\varepsilon_\infty \qquad (7.10)$$

and, by the law of total energy conservation it equals to the quantum motion nonlocality energy of a particle at infinity. Expressions (7.5) and (7.10) make it possible to find unknown values $r_2$, $\tau_0$, $\delta\varepsilon_\infty$ and calculate $D$ through the given kinetic energy of the particle at infinity.

If in (7.8) we assume that $\beta^2(r) \gg 4\alpha^2$ in the whole area of $r_2 - r_0$ (though $\beta(r_1)=0$) and take into account that $\beta$ changes its sign in integration ranges, a known quasiclassical result of the barrier transparency as a function of the particle energy obtained by G. Gamow, R. Gourney and E. Condon follows from (7.9) [4]. However, the total energy of the particle here should be substituted in the form of $E = E_k + \delta\varepsilon_\infty$. Previously for the better agreement with the experiment some additives to $E_k$, caused by decreasing the Coulomb barrier of the nucleus by surrounding electrons, were calculated [3]. The shielding effect is very difficult to measure experimentally. In theory, the effect evaluations differed with different authors due to a variety of approaches to the multielectron problem. In our case the $\delta\varepsilon_\infty$ has a fundamentally different meaning, this is a quantum motion nonlocality energy carried away from a nucleus which can be measured.

Now we can evaluate the effect connected with the presence of $\delta\varepsilon_\infty$. Solution (7.8) ignores coordinate derivatives of $\varphi$, which are necessary for finding point $r_2$, and is invalid. Let us assume that in the first approximation a quantum particle in infinite motion moves in the nucleus field with a classical momentum value, then the quantum nonlocality energy remains unchanged. Using equations (7.5), (7.10), we obtain:

$$\delta\varepsilon_\infty / E_k \leq (8Z^2 e^4 m/(\pi^2 \hbar^2 E_k)+1)^{1/2} - Ze^2(8m)^{1/2}/(\pi \hbar E_k^{1/2}) \qquad (7.11)$$

For example, at alpha-decay of $^{210}Po$ (Z=84-2) with the kinetic energy of particles of 5.3 MeV $\delta\varepsilon_\infty \leq 178$ keV. The relative excessive heating is 3.3%. In accordance with [3] the empirical amendment will be 131.6 keV in this case and it decreases with the nucleus charge. The value of $\delta\varepsilon_\infty$ can be calculated more precisely if necessary.

4. V.K. Nevolin. On Heat Emission of Alpha-sources. Ingenernaya Phizika Journal. 2009. No.3. P.10.
5. V.K. Nevolin. Thermal Effect on Anode at Field Emission. JPL Letters. Technical Physics Letters, 2006, V.32, No. 12, pp. 1030-1032.
.

## 8. Measuring Energy of Quantum Particles in Infinite Motion [9]

To measure any classical physical quantity one most relevant method is commonly used. Duplicate measuring methods are not used, as a rule. The things with measuring quantum physical quantities are quite different. There often is a vital necessity to measure the same quantity by at least two different ways. As it is shown below, realization of these methods using nanotechnologies [1] results in discovering new regularities and developing new devices based on them.

To measure the quantum nonlocality energy of particles in infinite motion (these can be electrons, protons, neutrons, alpha particles and other quantum particles) well-known methods of measuring quantum particle energy are used in each particular case, for example, mass-spectrometry [2], calorimetry [3] and others. These measuring methods are prototypes of the proposed means of measurement.

Disadvantages of known methods of measuring the energy of quantum particles in infinite motion feature are as follows: if the methods allow the same energy to be measured with an error inherent to each method (the one which suits is chosen from them), there is no vital necessity to measure it using at least two different methods. This approach is unsuitable for measuring the energy of quantum particle infinite motion, as the total energy of particles consists of two different types of energy, the energy of classical translation movement with some mean momentum value and pure quantum energy of motion. If measurements of this kind had been carried out earlier and, namely, the energy of particles in infinite motion in the free space had been measured using, for example, the calorimetry method and the energy of the same particles had been defined, for example, by the mass-spectrometry method, the difference between these energies would have made it possible to discover and measure the energy of quantum motion nonlocality earlier. The total energy of quantum particles moving in a stationary external field is an invariant of motion and can be presented as follows [4, 5]:

$$E = E_k(\vec{r}) + U(\vec{r}) + \Delta\varepsilon(\vec{r}), \qquad (8.1)$$

where $E_\kappa$ is the mean kinetic energy of a particle, $U$ is the potential energy of an external field, $\Delta\varepsilon$ is the mean energy of quantum particle nonlocality, $\vec{r}$ - spatial coordinates. In the free space $U(\vec{r}) = 0$ and the values of $E_\kappa$ and $\Delta\varepsilon$ remain constant.

Our aim is to measure the energy of quantum nonlocality of particles in infinite motion $\Delta\varepsilon$.

While measuring $\Delta\varepsilon$ in the external field one must know additionally its value and spatial distribution. Hence, it is most convenient to take these measurements in the free space when the potential energy equals zero. This can be achieved by measuring the total energy of each type of particles by their full stopping in the environment with given properties and measuring calorific effect of heat release, for example, by means of microthermocouples. Then the kinetic energy of these particles is measured. If they carry a charge and their masses are known the measurements are fulfilled by mass-spectrometry (the particle trajectory curvature radius is measured in a transverse magnetic field), if particles are not charged the recoil momentum of elastic scattering on the target made from a relevant material placed on a torsion balance is measured [6]. In all cases it is necessary to have a counter for particles per unit of time which are to be measured. If the particles are charged the value of their current is measured, if the current is very small a galvanometer is used, for example. If particles do not carry any charge, for example, neutrons, the particle flow is measured with radiometers.

Let us consider a measurement method by the example of electron tunneling [4].

At electron tunneling from the cathode through a triangular barrier at field emission, electrons carry Fermi energy away from the cathode, that is, the energy of quantum motion nonlocality, gain kinetic energy in the external field between electrodes and deliver it to the anode. If the Fermi energy of electrons in the cathode exceeds the Fermi energy in the anode some additional heat release is possible in comparison with the classical case in the anode, and anode underheating at the reverse ratio of Fermi energies [1]. This is an analog to Peltier effect having, however, one fundamental difference: tunneling electrons carry Fermi energy from the cathode and transfer it in the space between electrodes. This effect was not observed previously due to specific character of earlier experiments carried out at large interelectrode voltages when the effect is vanishingly small (see formula (2) below).

The relative heat emission in the anode as a function of the voltage applied will obey the following law:

$$\Delta Q/Q = 1 + (\Delta\varepsilon_{f1} - \Delta\varepsilon_{f2})/eU, \qquad (8.2)$$

where $Q = IU$, $U$ is the voltage applied, $I$ is the current in the circuit, $\Delta\varepsilon_{f1}, \Delta\varepsilon_{f2}$ are Fermi energies of the cathode and the anode. It follows from (8.1) that anode overheating or underheating is possible depending on the relation between Fermi energies of electrodes, heat release being classical only in the specific case of similar electrodes. Generally, there are two unknown quantities $\Delta\varepsilon_{f1}, \Delta\varepsilon_{f2}$ in formula (1) and to measure one of them, the reference electrode, for example anode, is necessary. If the materials the electrodes are made of are similar, the kinetic energy of electrons coming to the anode can be defined by means of the thermal effect taking into account initial thermal velocities [5]. As the Fermi energy of electrodes can be measured by other method as well [7], the given example can be looked at as another method of measuring Fermi energy. To obtain the field emission mode (a triangular barrier) the voltage applied must exceed the largest work function of the cathode and anode $eU > \max(e\varphi_1, e\varphi_2)$. However, the voltage applied should not be too high, otherwise the effect will be vanishingly small.

Let us consider the measuring method by the example of alpha-decay.

In alpha-decay charged particles must carry away both kinetic energy and quantum motion nonlocality energy which can be measured as a difference between the total energy of particles thermalized in the environment with given properties, and the kinetic energy of incoming particles. The presence of a charge allows the current and the kinetic energy of particles to be measured in a transverse magnetic field by means of mass-spectrometry. Thus if the kinetic energy of outgoing particles is known in alpha-sources, and the heat emission in the environment with specified properties is measured, the amount of heat should be somewhat bigger due to additional thermalization of the quantum motion nonlocality energy. Actually, this effect does take place [8], however, it has not been measured directly. The kinetic energy of alpha-particles is measured with mass-spectrometers precisely and to prove their tunneling origin the results were compared to theoretical formulae of tunneling. It was found that there was some deficit of kinetic energy equaling $\Delta\varepsilon$. Essentially, to agree with the theory the empiric formula was made up [8]:

$$\Delta\varepsilon = 73Z^{4/3} + 65Z^{5/3}, \text{эВ}, \qquad (8.3)$$

where $Z$ the number of charges of the daughter nucleus. For example, at $\alpha$ - decay of $^{210}Po$ (Z=84-2) with the kinetic energy of particles of 5.3 MeV the empirical correction to $E_k$ makes 131.6 keV for this case. Relative "excessive" heating of the target must be 3.3% which can be measured with state-of-the-art devices. The natural alpha-particle radiation linewidth measured with a mass spectrometer is about a few millielectronvolts.

The calorimetric method of measuring is less precise in comparison with mass-spectrometry.

To increase the accuracy of measuring the quantum motion nonlocality energy a differential method of measuring the total energy of particles is offered. In some area on the way of alpha particles a constant electric field is placed which is characterized by specified potential difference, for example, slowing down. Then, the following expression for the total energy of a particle in accordance with formula (8.1) will be obtained:

$$E_1 = E_k - qV + \Delta\varepsilon \tag{8.4}$$

where $q$ is the charge of a particle, $V$ is the potential difference which the particle passes trough, $E_k$ is the kinetic energy of the particle. The energy of particles is measured by a calorimetric method. Then the sign of the electric field is changed, the expression for the total energy of the particle will be:

$$E_2 = E_k + qV + \Delta\varepsilon \tag{8.5}$$

and the total energy of the particle is measured by a calorimetric method. Then the difference between the two measurements must be equal to:

$$\Delta E = E_2 - E_1 = 2qV \tag{8.6}$$

With some decrease in the voltage applied this difference will vanish. It is this voltage value that characterizes the error of the method. The half-sum of the measured values according to (8.5) and (8.6) will give the total energy to define subsequently the quantum motion nonlocality energy of particles.

While designing and exploiting precision alpha-sources of heat it is necessary to consider the fact that the emission of thermal energy at alpha-decay is somewhat higher than it can be expected based on kinetic energy measured experimentally with mass spectrometers (with high precision as a rule).

The energy of quantum nonlocality of particle infinite motion as mentioned above can be used in developing refrigerating devices [1] based on nanotechnology.

Reference
1. V.K. Nevolin. Heat Emission or Absorption Device. RF Patent with priority of 28.08.2008.
2. A.A. Sysoev, M.S. Chupakhin. Introduction into Mass-spectrometry. M. 1963.
3. E. Calve, A. Prat. Microcalorimetry. Translation from French. M. 1963.
4. V.K. Nevolin. Thermal Effect on Anode at Field Emission. JPL Letters. Technical Physics Letters, 2006, V.32, No. 12, pp. 1030-1032.

## 9. Quantum Statistic Resonance at Electron Beam Interaction with Laser Radiation [1]

Let us consider interaction between a delicately divergent monochromic electron beam and an opposite single-mode laser beam, Fig. 1.

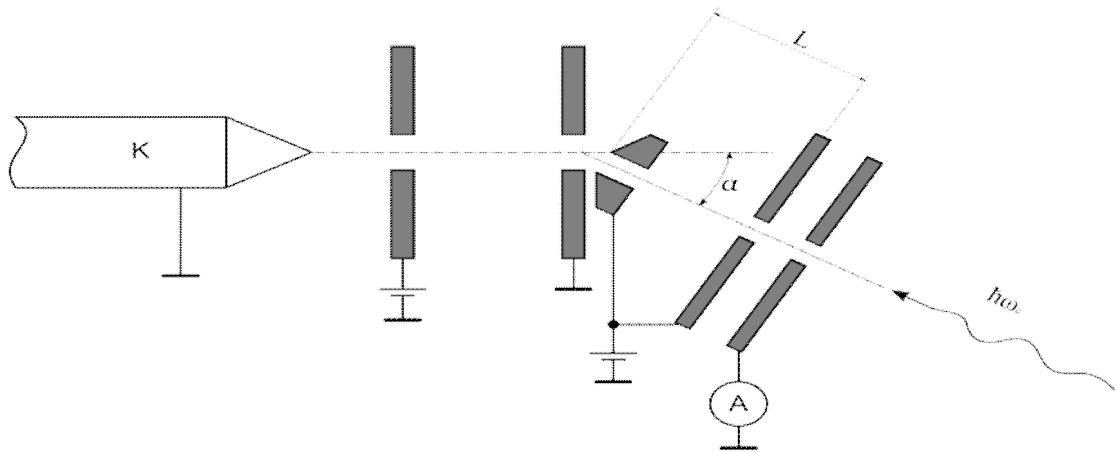

Figure 1.

Schematic diagram of a potential experimental facility:

C-cathode; L-distance between electrodes: A-galvanometer current

Electrons are extracted from the cathode by means of field emission with Fermi momentum $k_f$, thus $k_f^2 = k_t^2 + k_l^2$, and the transverse component of momentum is small compared with the momentum directed along the beam $k_t/k_l \ll 1$. The electron beam is so rarified that the interaction between electrons might be ignored. Electrons move steadily in the free space between the two electrodes with collimating apertures, with a collecting electrode behind the last of them. When light interacts with an electron beam the Compton effect is possible which results in light scattering the electron beam and reduced current to the collecting electrode. The scattering characteristic here will be monotonous functions versus change in the energy of the electron beam in the free inter electrode space between. The electric current on the collecting electrode after the collimating aperture will change monotonously.

The effect of random interaction between the transverse electric field of the light wave and the electric charge, which changes the transverse component of electron momentum. However, in this case time-resonant interaction between the transverse electric field of the light wave and the electric beam, if (see Formula (2.9)):

$$\omega_c = 2\omega = 2\vec{k}\cdot\vec{p}/m = 2\cos\alpha \cdot k \cdot p/m \qquad (9.1)$$

where $\omega_c$ is a cyclic frequency of light, $\alpha$ is the angle between vectors. Factor 2 takes into account the fact that harmonic oscillations of the probability density in accordance with Formula (2.6) take place at doubled frequency $\omega$. The transverse component of the wave vector $\vec{k}$ due to statistics is equiprobably distributed over all directions, which is required from the polarization vector of the light wave. When interacting waves have temporal synchronism the value of the light wave electric field will be nonuniform along the electron beam. However, the wave vector of the light wave $k_c$ is much larger than the Fermi wave vector $k_f$, $k_c \gg k_f$, and the laser wavelength will "cover" a number of oscillations of probability density. Thus, excitation of the transversal component of the motion energy does not require oscillation phase synchronism in the probability density of particles participating in motion, it is essential that they oscillate at the same frequency. Then, in the first approximation for the effective force $F(t)$, the average value over half period can be used:

$$F(t) = \frac{2eD}{\pi}\sin(\omega_c \cdot t + \beta)$$

where $D$ is the amplitude of light wave electric field, $e$ is the charge of an electron.

The light wave electric field will build up the transversal component of electron momentum increasing the transversal component of the beam energy, the longitudinal component of the electron beam energy and the frequency of spatial probability density oscillations remaining unchanged. In the case of resonance the incoming current will drastically decrease on the collecting electrode after the collimating electrode. The effect value depends on the duration of resonant interaction between light and the beam.

The forced oscillations energy which is gained by the electron in the light wave field for the time period T can be estimated by the equation:

$$\varepsilon_c = \frac{1}{2m}\left|\int_0^T F(t)e^{-i2\omega t}dt\right|^2$$

We have the following in the case of resonance:

$$\varepsilon_c = \frac{e^2 D^2}{2\pi^2 m \omega_c^2} \left[(\omega_c T)^2 + \sin^2 \omega_c T + (\omega_c T/2)(\sin(\omega_c T + \beta) - \sin \beta)\right] \quad (9.2)$$

Let us assess the main parameters necessary to provide resonance excitation of the quantum component of electric motion energy. If the gap between electrodes with collimating apertures equals $L$, where electrons move freely at a speed $v$, then $T=L/v$. Electrons enter the free space having passed the accelerating potential difference $U$, $v = (2eU/m)^{1/2}$. We ignore the thermal spread of electrons over energies compared to the Fermi energy and the energy which electrons gain in the electric field. We consider the electrons to tunnel mainly from top Fermi surface (transversal components of quasi-momentum are small). Hence, $k_f = (2m\varepsilon_f/\hbar^2)^{1/2}$ and the resonant frequency of laser radiation equals:

$$\omega_c = 4\cos\alpha (eU \cdot \varepsilon_f / \hbar^2)^{1/2} \quad (9.3)$$

Colliding beams should not be coaxial for the light beam after passing collimating apertures not to cause the photoelectric emission from the cathode, $\alpha > 0$. For the silver cathode with the Fermi energy $\varepsilon_f = 5.5$ eV and the translation motion energy of electrons $eU = 1$ eV the maximum energy of a light quanta must be $\hbar \cdot \omega_c = 9.4$ eV. This is the ultraviolet range of lasers. For lower energy of laser radiation $\alpha \to \pi/2$, however, the intensity of the gathered electron beam will considerably decrease in this case. Fig. 1 shows the schematic drawing of the experimental facility.

The value $\omega_c T$ in Formula (7) equals $\omega_c T = 2k_f L \gg 1$, and it can be seen that the initial values of light wave phases $\beta$ are not significant in the resonance, and for the purpose of estimation we restrict ourselves to the first term in this formula. Thus, we have:

$$\varepsilon_c = \frac{e^2 D^2 T^2}{2\pi^2 m} \quad (9.4)$$

The amplitude of the light wave electric field $D$ depends on the intensity of the laser beam in a known manner.

In conclusion it should be noted that the experimental proof of "cold" excitation of quantum energy component for particles in infinite motion by laser radiation is of fundamental importance for solving applied problems, and will be another evidence of the wave character of quantum particle motion.

Reference
1. V.K. Nevolin. Quantum Statistic Resonance at Electron Beam Interaction with Laser Radiation. Prikladnaya Fizika Journal. 2011. No. 3. P.73-76.

## 10. Particle Motion in Potential Step Field

Let us consider one-dimensional stationary motion of particles having energy $E$, momentum $P_1$ in the field of a rectangular barrier of height $U_0$, occupying the right half space ($E>U_0$), see Fig. 1.

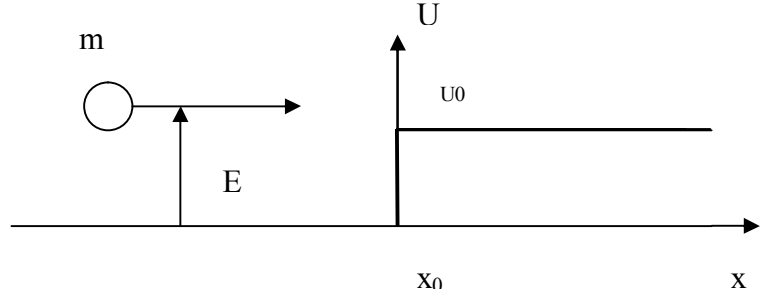

Fig. 1

It is a model problem of quantum mechanics. While in classical mechanics the reflection coefficient of particles passing over the step is equal to zero, this coefficient in quantum mechanics differs from zero. Actually, this coefficient can be easily calculated (see, for example [1]), however, it turns out not to depend on Plank constant, in other words, the quantum effect does not depend on the quantum constant. In the text book [1] it is called an illusory contradiction and attributed to the fact that the de Broglie wavelength in this kind of problem is always not less than the area of the potential jump which is equal to zero. This is unsatisfying explanation. In the following solved problem with a rectangular barrier transmission and reflection coefficients do depend on Plank constant, moreover there is a purely quantum resonance effect of passing the particle over the barrier when $D=1$. Evidently, it is not potential jumps that matter, but the existing way of describing infinite motion of quantum particles. We shall illustrate that such contradiction does not occur in quasi-hydrodynamic presentation of quantum particle motion.

In the left half space there are two stationary opposite particle flows. In accordance with (3.7) and (3.8) the following system of equations should be solved in the left half-space:

$$\frac{P_i^2}{2m} - \frac{\hbar^2}{4m\rho_{13}}\frac{\partial^2 \rho_{13}}{\partial x_i^2} + \frac{\hbar^2}{8m\rho_{13}^2}\left(\frac{\partial \rho_{13}}{\partial x_i}\right)^2 = E \quad (10.1)$$

$$i=1, 3$$

$$m\frac{\partial \rho_{13}}{\partial t} + P_1\frac{\partial \rho_{13}}{\partial x_1} - P_3\frac{\partial \rho_{13}}{\partial x_3} = 0 \quad (10.2)$$

Here $i=1$ corresponds to the incident particle, $i=3$ corresponds to the reflected particle, $\rho_{13}$ – is the probability density of particles in the left half-space. We seek for the solution in the following form:

$$\rho_{13} = \rho_1(tP_1/m\,\delta x_1 - x_1/\delta x_1) \cdot \rho_3(tP_3/m\,\delta x_3 + x_3/\delta x_3)$$

And obtain:

$$\rho_1 = \rho_{10} \cos^2 \pi(tP_1/m\,\delta x_1 - x/\delta x_1)$$

$$\rho_3 = \rho_{30} \cdot \cos^2(\pi(tP_3/m\,\delta x_3 + x_3/\delta x_3) + \phi_{03})$$

where $\dfrac{\pi}{\delta x_i} = \dfrac{\delta P_{xi}}{\hbar}$. The initial phase of probability density of incident particles is taken equal to zero for convenience.

In the right half-space it is necessary to solve the following system of equations:

$$\frac{P_2^2}{2m} + U_0 - \frac{\hbar^2}{4m\rho_2}\frac{\partial^2 \rho_2}{\partial x^2} + \frac{\hbar^2}{8m\rho_2^2}\left(\frac{\partial \rho_2}{\partial x}\right)^2 = E \quad (10.3)$$

$$m\frac{\partial \rho_2}{\partial t} + P_2\frac{\partial \rho_2}{\partial x} = 0 \quad (10.4)$$

We obtain the solution to the probability density in the form of:

$$\rho_2 = \rho_{20} \cos^2(\pi(tP_2/m\,\delta x_2 - x_2/\delta x_2) + \phi_{20})$$

We shall illustrate boundary conditions of this problem by means of quantum equations of motion written in Schroedinger representation. The wave function (2.7) of a free particle is a superposition of de Broglie plane waves and, naturally, the solution to Schroedinger equation:

$$\Psi(\mathbf{r},t) = \sqrt{\rho_0}\cos\left(\frac{\delta\mathbf{p}(\mathbf{r}-t\mathbf{p}/m)}{\hbar}\right)e^{\frac{i(\mathbf{pr}-Et)}{\hbar}} \quad (10.5)$$

where

$$E = \mathbf{p}^2/2m + (\delta\mathbf{p})^2/2m$$

Let us calculate the density of particle momentum probability flow using Formula (3.3)

$$\mathbf{J} = \frac{1}{2m}(\Psi^*\hat{\mathbf{P}}\Psi + \Psi\hat{\mathbf{P}}^*\Psi^*) \quad (10.6)$$

Thus, we obtain

$$\mathbf{J} = \frac{\rho_0\mathbf{p}}{m}\cos^2\left(\frac{\delta\mathbf{p}(\mathbf{r}-t\mathbf{p}/m)}{\hbar}\right) = \rho(\mathbf{r},t)\cdot\mathbf{p}/m \quad (10.7)$$

Boundary conditions for wave functions are formulated on the mathematical grounds. We convert the boundary conditions for wave functions into the form having a physical meaning. In the simplest case of one-dimensional problem we have:

$$\Psi_a(t,x_0) = \Psi_b(t,x_0) \qquad \Psi_a^*(t,x_0) = \Psi_b^*(t,x_0) \quad (10.8)$$

Having multiplied these equations we obtain:

$$\rho_a(t,x_0) = \rho_b(t,x_0) \quad (10.9)$$

Boundary conditions (10.8) must fix the fact of reaching the boundary – collision with the wall i.e.

$$\rho_a(t_0, x_0) = \rho_b(t_0, x_0) \tag{10.10}$$

where $t_0$ is the time of reaching the boundary.

Other boundary conditions are:

$$\frac{\partial \Psi_a(t, x_0)}{\partial x} = \frac{\partial \Psi_b(t, x_0)}{\partial x} \qquad \frac{\partial \Psi_a^*(t, x_0)}{\partial x} = \frac{\partial \Psi_b^*(t, x_0)}{\partial x} \tag{10.11}$$

Multiplying the left and right parts of Equations (10.8) and (10.11) and subtracting we obtain

$$\Psi_a \frac{\partial \Psi_a^*(t, x_0)}{\partial x} - \Psi_a^* \frac{\partial \Psi_a(t, x_0)}{\partial x} = \Psi_b \frac{\partial \Psi_b^*(t, x_0)}{\partial x} - \Psi_b^* \frac{\partial \Psi_b(t, x_0)}{\partial x} \tag{10.12}$$

This equation expresses the equality of probability flows (see (10.6)) on the boundary with an accuracy to the constant coefficient

$$J_a(t, x_0) = J_b(t, x_0)$$

Flow densities of particle probability reach the boundary at the moment $t_0$

$$J_a(t_0, x_0) = J_b(t_0, x_0)$$

The formulae mentioned above show that the solution to the problem of particle motion in the potential step field must give the same result both when the problem is solved in quasi-hydrodynamic representation and in Schroedinger representation.

Probability flow densities and probability densities must remain continuous on the boundary of the rectangular step.

$$P_1 \rho_1(t_0, x_0) - P_3 \rho_3(t_0, x_0) = P_2 \rho_2(t_0, x_0),$$

$$\rho_1(t_0, x_0) + \rho_3(t_0, x_0) = \rho_2(t_0, x_0)$$

$$E = \frac{P_1^2}{2m} + \frac{\pi^2 \hbar^2}{2m (\delta x_1)^2}$$

$$E = \frac{P_3^2}{2m} + \frac{\pi^2 \hbar^2}{2m (\delta x_3)^2}$$

$$E = \frac{P_2^2}{2m} + U_0 + \frac{\pi^2 \hbar^2}{2m (\delta x_2)^2}$$

Here $x_0$ is the coordinate of the rectangular barrier wall. An incident particle reaches the wall at the moment $t_0 = x_0 m / P_1$. If the particle is reflected with the amplitude of probability density $\rho_{30}$ it carries away the information of this fact in the point $x_0, t_0$ by means of the phase

$$\phi_{30} = \pi(n - t_0 P_3 / m \delta x_3 - x_0 / \delta x_3), n = 0,1,2,3... \qquad (10.13)$$

If the particle overcomes the rectangular barrier with the amplitude of probability density $\rho_{20}$, it carries away the information of this fact in the point $x_0, t_0$ by means of the phase

$$\phi_{20} = \pi(n - t_0 P_2 / m \delta x_2 + x_0 / \delta x_2), n = 0,1,2,3... \qquad (10.14)$$

Then the boundary conditions can be rewritten in the form:

$$P_1 \rho_{10} - P_3 \rho_{30} = P_2 \rho_{20} \qquad (10.15)$$

$$\rho_{10} + \rho_{30} = \rho_{20} \qquad (10.16)$$

For the transmission coefficient $D$ we have:

$$D = \frac{P_2(P_1 + P_3)}{P_1(P_2 + P_3)} \qquad (10.17)$$

Let us normalize the obtained solutions for $\rho_1, \rho_2, \rho_3$ within one oscillation of the probability density, we will have:

$$\delta x_1 = 2/\rho_{10}, \quad \delta x_2 = 2/\rho_{20}, \delta x_3 = 2/\rho_{30}$$

Then the expression for the particle energy can be written as follows:

$$E = \frac{P_1^2}{2m} + \frac{\pi^2 \hbar^2 (\rho_{10})^2}{8m}$$

$$E = \frac{P_3^2}{2m} + \frac{\pi^2 \hbar^2 (\rho_{30})^2}{8m}$$

$$E = \frac{P_2^2}{2m} + U_0 + \frac{\pi^2 \hbar^2 (\rho_{20})^2}{8m}$$

To calculate the barrier transparency coefficient (10.17), we find expressions for unknown momenta $P_2, P_3$ using the written values of the total energy and the definition for $D = \rho_{20} P_2 / \rho_{10} P_1$.

$$P_3 = \left[ mE - \sqrt{m^2 E^2 - \pi^2 \hbar^2 (1-D)^2 \rho_{10}^2 P_1^2 / 4} \right]^{1/2} \qquad (10.18)$$

$$P_2 = \left[ m(E - U_0) + \sqrt{m^2(E - U_0)^2 - \pi^2 \hbar^2 D^2 \rho_{10}^2 P_1^2 / 4} \right]^{1/2} \qquad (10.19)$$

Here the expressions for moment have been obtained in an inexplicit form through the transmission coefficient $D$. Formulae (10.17) – (10.19) together make it possible to calculate the barrier transmission coefficient in an explicit form. Values $E$, $P_1$, $\rho_{10}$ are considered known for particles incident onto the potential step. In this case $E = \dfrac{P_1^2}{2m} + \dfrac{\pi^2 \hbar^2 (\rho_{10})^2}{8m}$.

The expressions obtained for moment provide the classical limit $D=1$ when $\hbar \to 0$. The transmission coefficient $D = 0$ with the total energy value $E=U_0$ in this case $P_3=P_1$.

As a rule, the quantum component of particles incident onto the barrier is small i.e.

$$\delta \varepsilon_0 = \frac{\pi^2 \hbar^2 (\rho_{10})^2}{8m} \ll \frac{P_1^2}{2m} = E_k$$

Hence, it follows from (10.18) (10.19) that:

$$P_3 = P_1(1-D)\sqrt{\delta \varepsilon_0 / E_k}, \qquad P_2 = \sqrt{2m(E_k - U_0)} \qquad P_1 = \sqrt{2mE_k} \qquad (10.20)$$

i.e. reflected momenta values make a small fraction of incident moment values.

Formulae obtained for the transmission coefficient $D$ differ substantially from the solution to the similar problem for quantum particles incident onto the potential step, where the amplitude of probability density is constant over the whole space (see [1]).

Let us show what makes the traditional solution to this problem different from the solution given above.

To solve Equation (10.1) we shall assume that there is a superposition of quantum states of incident and reflected particles, i.e. their coordinates are indistinguishable $x_1 = x_3 = x$

Let $\rho_1(x)$ and $\rho_3(x)$ be solutions to Equation (10.1). It can be shown that the superposition of these solutions in the form:

$$\rho_{13}(x) = \rho_1 + \rho_3 \pm 2\sqrt{\rho_1 \rho_3}$$

is also a solution to this equation, if $E = P_1^2/2m = P_2^2/2m$. However, for the coordinates of incident and reflected particles to be indistinguishable along the direction of motion they must be delocalized over the whole space. Let us fulfill the relevant transition in solutions for $\rho_1, \rho_3, \rho_2$ ($\delta x_i \to \infty$), then we obtain boundary conditions in the form:

$$\rho_{10} + \rho_{30} + 2\sqrt{\rho_{10}\rho_{30}} = \rho_{20}$$

$$P_1(\rho_{10} - \rho_{30}) = P_2 \rho_{20}$$

which provides obtaining formulae of Problem 1 for $R$ and $D$ from [6], which are independent from the Plank constant and, therefore, do not provide the classical limit when $\hbar \to 0$.

$$D = 4P_1 P_2 / (P_1 + P_2)^2$$

The existence of classical limit for the transmission coefficient *D* should not certainly depend on the steepness of the potential step boundary. The reason is in the traditional view of a free quantum particle as de Broglie wave.

## 11. Tunneling

The tunneling problem was solved in quasi-hydrodynamic representation in Paragraph 7, when alpha-decay of nuclei was studied. Tunneling was looked at from a potential well with finite motion of particles with a zero translation motion component. We shall consider particle tunneling in a more general case when particles collide with a rectangular wall with some momentum.

Let us consider a particle passing over a rectangular barrier of a finite height $U_0$ ($E < U_0$) and finite width *a*, Fig.1.

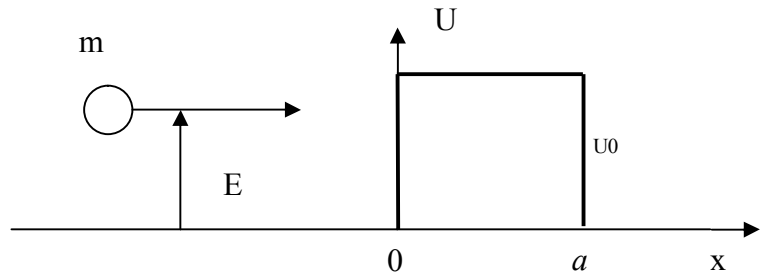

Fig.1

As there is no macroscopic particle momentum in the vicinity of the barrier, it is necessary to solve a system of Equations (3.2), (3.6).

$$m \frac{\partial \rho}{\partial t} + div \, \mathbf{J} = 0, \quad (3.2)$$

$$\frac{\partial (\mathbf{J}/\rho)}{\partial t} = -\nabla \left( \frac{J^2}{2m\rho^2} + U + \frac{\hbar^2 (\nabla \rho)^2}{8m\rho^2} - \frac{\hbar^2 \Delta \rho}{4m\rho} \right), \quad (3.6)$$

We shall make some simplifying assumptions to solve the system of equations. We assume that $\frac{\partial}{\partial t}(\frac{J}{\rho}) = 0$ at tunneling in stationary potential fields, i.e. the energy of tunneling particles remains constant

$$E = \frac{J^2}{2m\rho^2} + U_0 - \frac{\hbar^2}{4m\rho}\Delta\rho + \frac{\hbar^2}{8m\rho^2}(\nabla\rho)^2 = const \quad (11.1)$$

To simplify the formulae further we shall consider one-dimensional spatially localized particles – needle states. If there are transversal components of quantum oscillation energy, they are "pulled" over the barrier without unchanged as they are motion invariants. We shall take into account the transversal components in final formulae.

Then, it follows from (11.1) that $J/\rho$ depends only on coordinate $x$ and

$$\frac{J}{\rho} = \sqrt{\frac{\hbar^2}{2\rho}\frac{\partial^2\rho}{\partial x^2} - \frac{\hbar^2}{4\rho^2}(\frac{\partial\rho}{\partial x})^2 - (U_0 - E)2m} \quad (11.2)$$

To fulfill this relation it is necessary to accept $\rho = \rho_t(t) \cdot \rho_x(x)$.

Then, Equation (3.2) can be solved by the method of variable separation in quadratures. We obtain

$$J(x,t) = \exp(t/\tau)(J_0 - \frac{m}{\tau}\int_0^x \rho_x(x)dx) \quad (11.3)$$

Here $\tau = m\hbar/P_1^2$, $x=0$ is the position of the front wall of the barrier, where the flow density of particle probability equals $J_0$. The solution of Equation (3.2) can be written in a different form

$$J(x,t) = J_0 \exp(\frac{1}{\tau}(t - m\int_0^x \frac{\rho_x(x)}{J_x(x)}dx)), J_x(x) = J_0 - \frac{m}{\tau}\int_0^x \rho_x(x)dx, \quad (11.4)$$

One can make sure that these are similar solutions by differentiating them with respect to $x$ coordinate. It follows from (11.4) that the probability density flow $J(x,t) = J_0$ is transferred in accordance with the law

$$t = m\int_0^x \frac{\rho_x(x)}{J_x(x)}dx \quad (11.5)$$

Having substituted Solution (11.3) into (11.1) we obtain a system of equations:

$$E = U_0 + \frac{J_x^2(x)}{2m\rho_x^2(x)} + \frac{\hbar^2}{8m\rho_x^2}(\frac{d\rho_x}{dx})^2 - \frac{\hbar^2}{4m\rho_x}\frac{d^2\rho_x}{dx^2}$$

$$J_x(x) = J_0 - \frac{m}{\tau}\int_0^x \rho_x(x)dx \quad (11.6)$$

We can write this system in a different way:

$$2J_x'''J_x' - (J_x'')^2 - \chi^2(J_x')^2 - \theta^2 J_x^2 = 0, \rho_x(x) = -\frac{\tau}{m}J_x'$$

where $\chi^2 = \frac{8m}{\hbar^2}(U_0 - E), \theta^2 = 4P_1^4/\hbar^4$, the prime mark indicates x-derivatives. The solution $J_x$ has been found in the form:

$$J_x(x) = A\exp(\pm\beta x)$$

where $\beta = \left[\chi^2/2 + \sqrt{\chi^4/4 + \theta^2}\right]^{1/2}$, $A$ is constant.

Then, we calculate the tunneling barrier-transmission coefficient $D$. For this purpose one can use solutions for probability density obtained earlier in areas before the barrier and after passing the barrier. We assign index 1, as previously, to particles incident onto the barrier, and index 3 to reflected particles, index 2 is given to particles tunneling in the vicinity of the barrier, and index 4 – to particles which have passed the barrier. Thus, we have:

$$\rho_1 = \rho_{10}\cos^2\pi(tP_1/m\delta x_1 - x/\delta x_1)$$

$$\rho_3 = \rho_{30}\cdot\cos^2(\pi(tP_3/m\delta x_3 + x/\delta x_3) + \phi_{03})$$

$$\rho_2(x,t) = \rho_{20}\exp(\frac{t}{\tau} - \beta x), J_2(x,t) = J_{20}\exp(\frac{t}{\tau} - \beta x)$$

$$\rho_4 = \rho_{40}\cos^2(\pi(tP_4/m\delta x_4 - x/\delta x_4) + \phi_{04}).$$

Laws of energy conservation for free particles can be represented in the following way:

$$E = \frac{P_1^2}{2m} + \frac{\rho_{10}^2 \pi^2 \hbar^2}{8m} + \varepsilon_\perp$$

$$E = \frac{P_3^2}{2m} + \frac{\rho_{30}^2 \pi^2 \hbar^2}{8m} + \varepsilon_\perp$$

$$E = \frac{P_4^2}{2m} + \frac{\rho_{40}^2 \pi^2 \hbar^2}{8m} + \varepsilon_\perp$$

Boundary conditions can be written, as previously, in this form:

$$P_1\rho_1(t_0,x_0) - P_3\rho_3(t_0,x_0) = J_2(t_0,x_0) \qquad J_2(t_0,x_0+a) = P_4\rho_4(t_0,x_0+a)$$

$$\rho_1(t_0,x_0) + \rho_3(t_0,x_0) = \rho_2(t_0,x_0) \qquad \rho_2(t_0,x_0+a) = \rho_4(t_0,x_0+a)$$

We assume further, $x_0 = 0, t_0 = 0$, as the front wall of the barrier is tied to coordinate $x_0=0$ in the solution for tunneling particles.

Solving the written system of equations together with relations for laws of energy conservation of free particles, one can calculate $D$.

$$D = \frac{P_4 \rho_4(a)}{P_1 \rho_{10}} \tag{11.7}$$

It follows from boundary conditions:

$$\rho_{4(a)} = \rho_2(a) = \rho_{20} e^{-\beta a} = (\rho_{10} + \rho_{30})e^{-\beta a}$$

Then $D$ equals:

$$D = \frac{P_4(\rho_{10}+\rho_{30})}{P_1 \rho_{10}} \exp(-\beta a)$$

At $a \to 0$ $D \to 1$, $\rho_{30} \to 0$ there must be $P_4 = P_1$. Then it follows from the energy conservation law: $\rho_{40} = \rho_{10}$. From the solution for probability density $\rho_4(a) = \rho_{10} \cos\varphi_4^2$.

Let us write out final formulae to calculate the tunnel barrier-transmission coefficient $D$.

$$D = \frac{\rho_{10} + \rho_{30}}{\rho_{10}} \exp(-a\beta) = \cos\phi_4^2 \tag{11.8}$$

It can be seen from (11.8) that the particles which have tunneled through the barrier preserve their energy, momentum, and the wave of probability density changes the phase.

From the law of conservation of energy of reflected particles and the definition $R = P_3 \rho_{30} / P_1 \rho_{10}$ we can find the expression for $\rho_{30}$

$$\rho_{30} = \frac{1}{\gamma\sqrt{2}} \left[ 2m(E-\varepsilon_\perp) - \sqrt{4m^2(E-\varepsilon_\perp)^2 - 2\gamma^2(1-D)^2 P_1^2 \rho_{10}^2} \right]^{1/2} \tag{11.9}$$

$$\beta = \chi\left[1/2 + \sqrt{1/4 + \theta^2/\chi^4}\right]^{1/2}$$

$$\gamma = \pi \hbar / 2$$

$$\chi^2 = \frac{8m}{\hbar^2}(U_0 - E), \theta^2 = 4P_1^4/\hbar^4$$

Using Equations (11.8) and (11.9) one can calculate the barrier transmission ratio, if $E$, $P_1$, $\rho_{10}$ are given. The barrier transmission in the classical limit must be equal to zero, indeed, if we formally fix $\hbar \to 0$, we obtain $\beta \to \infty$ and $D \to 0$.

The conventional formula for the barrier under consideration (see [1], p.104) is as follows:

$$D_0 = \frac{4k_1^2 \chi_0^2}{(k_1^2 + \chi_0^2)^2 sh^2 a\chi_0 + 4k_1^2 \chi_0^2}, k_1 = P_1/\hbar \quad \chi_0 = (2m(U_0 - E_k)/\hbar^2)^{1/2} \tag{11.10}$$

Here $E_k = \frac{P_1^2}{2m}$ is the kinetic energy of a particle. Comparing Equations (11.10) and (11.8), (11.9), one can see that the tunneling of spatially localized particles differs from conventional tunneling. The preexponential factor in Equation (11.8) changes from 1 to 2. The exponent index $\beta$ includes parameter $\chi$ which will be written in the form convenient for comparison:

$$\chi = 2\sqrt{2m(U_0 - E_k - \varepsilon_\perp - \frac{p_{10}^2}{8m}\pi^2\hbar^2)}$$

It is evident that $\chi < 2\chi_0$, however in the formula for $\beta$ there is a parameter $\theta$, which is equal to

$$\frac{\theta^2}{\chi^4} = \frac{E_k^2}{4(U_0 - E)}$$

and which increases the coefficient $\beta$ to some extent. Thus, barrier-transmission coefficients should be compared for particular cases. Let us compare the formulae for the case when $a\chi_0 \gg 1$ as usual. Then, from (11.10) we have

$$D_0 = \frac{16k_1^2\chi_0^2}{(k_1^2 + \chi_0^2)^2} e^{-2a\chi_0} \qquad (11.11)$$

We shall carry out numerical comparison of transmission coefficients. We consider needle states, when $\varepsilon_\perp = 0$ and $\theta^2/\chi^4 \ll 1$, then $\beta = \chi = 2\chi_0$ and exponent indices coincide in formulae for $D$ and $D_0$. The difference in barrier-transmission coefficients will be defined by preexponential factors ratio. Let $k_1^2/\chi_0^2 = E_k/(U_0 - E_k) = 1$, then $D/D_0 \approx 0,5 < 1$. The situation changes radically with $P_1 = k_1\hbar \to 0$ when tunneling is possible mainly due to the energy of quantum particle fluctuations which are not considered by the conventional approach, then

$$D/D_0 \approx \chi_0^2/8k_1^2 \gg 1$$

Thus, tunnel transmission coefficients for spatially localized particles actually always differ from those calculated by means of traditional formulae.

The time of tunneling for spatially localized particles is of quite a definite value and, in accordance with formula (11.5) and $\tau = m\hbar/P_1^2$, equals:

$$\Delta t = \tau a\beta = \tau_0 \hbar \beta / P_1 , \qquad (11.12)$$

where $\tau_0$ is the transit time for the $a$-wide barrier, with the momentum $P_1$ $\tau_0 = \frac{am}{P_1}$

It can be seen that the process of tunneling can be slow when the momentum of a particle incident onto the barrier is small.

Reference

1. L.D. Landau, E.M. Lifshitz. Quantum Mechanics. Nonrelativistic Theory. M.: Nauka 1974. P.103.

**Conclusion**

Quasi-hydrodynamic representation of infinite motion of quantum particles with physical variables makes it possible to obtain new solutions which change the idea of them. The "cost" of the physical form of the initial system of equations is nonlinearity of one of them. For theoretical physics analytical solutions to quantum problems are invaluable. However, in the era of computer technologies solving quantum problems in quasi-hydrodynamic representation should not be of great difficulty, especially since transport problems in particular quantum devices are solved by means of computers due to their complexity.

Quantum particles in infinite motion beside classical kinetic energy always have quite certain quantum energy, so called, energy of quantum fluctuations. This approach eliminates contradictions of the traditional theory for transport phenomena, i.e. all the new quantum formulae and equations have a classical limit. In traditional quantum mechanics free particles are described by means of wave packets, as they are based on the notion of particle momentum fluctuations. However, the energy of these fluctuations which is transferred by free particles among other things was not taken into account by the energy conservation law.

The formulae for barrier-transmission coefficients have been obtained in the implicit form. Nominally, it is attributed to nonlinearity of one of the quantum motion equations. The process of transmitting barriers is essentially self-consistent for the probability density distribution, the statistic wave field of a particle changes with motion invariants preserved. In the mentioned traditional solutions to quantum problems this circumstance is not taken into account.

Considering spatial localization of free quantum particles gives, in our opinion, more correct and, sometimes, new relations for transport phenomena. Ultimately, using quasi-hydrodynamic representation is justified if there are new results which are or can be confirmed experimentally. In particular, experimental proof of resonant pumping the quantum component of the energy of infinite particle motion opens a fundamentally new approach to solving some applied problems and will provide an additional evidence of the wave character of infinite motion of quantum particles.

# Appendix 1
## Deriving Quantum Motion Equations in Quasi hydrodynamic Representation

Given below are initial Schroedinger equations necessary for further calculations.

$$i\hbar \frac{\partial \Psi}{\partial t} = \hat{H}\Psi \qquad -i\hbar \frac{\partial \Psi^*}{\partial t} = \hat{H}\Psi^* \qquad (Ap.\ 1.1)$$

Hamiltonian operator for a particle with mass *m*, which moves in an arbitrary potential field $U=U(r,t)$ has a form of:

$$\hat{H} = -\frac{\hbar^2}{2m}\Delta + U(r,t) \qquad (Ap.\ 1.2)$$

Multiplying the first Schroedinger equation by $\Psi^*$ and the other by $\Psi$ and subtracting the second equation from the first one, we obtain

$$i\hbar \frac{\partial \varrho}{\partial t} = \Psi^* \hat{H}\Psi - \Psi\hat{H}\Psi^* = -\frac{\hbar^2}{2m}(\Psi^*\Delta\Psi - \Psi\Delta\Psi^*) = -\frac{\hbar^2}{2m}div(\Psi^*\nabla\Psi - \Psi\nabla\Psi^*)$$

where $\varrho = \Psi\Psi^*$. Now we introduce probability flow density *J/m*

$$\frac{J}{m} = \frac{i\hbar}{2m}(\Psi\nabla\Psi^* - \Psi^*\nabla\Psi) \qquad (Ap.\ 1.3)$$

We get the law of particles mass conservation in the form:

$$m\frac{\partial \varrho}{\partial t} + div J = 0 \qquad (Ap.\ 1.4)$$

Then we calculate the derivative:

$$\frac{\partial\left(\frac{J}{\rho}\right)}{\partial t} = \frac{i\hbar}{2}\nabla \frac{\partial}{\partial t}(ln\Psi^* - ln\Psi) = \frac{i\hbar}{2}\nabla\left(\frac{1}{\Psi^*}\frac{\partial \Psi^*}{\partial t} - \frac{1}{\Psi}\frac{\partial \Psi}{\partial t}\right) = -\frac{1}{2}\nabla\left[\frac{\Psi\hat{H}\Psi^* + \Psi^*\hat{H}\Psi}{\varrho}\right]$$

Here Schroedinger equations given above are used. Substituting the explicit form of Hamiltonian operator we obtain:

$$\frac{\partial\left(\frac{J}{\rho}\right)}{\partial t} = -\nabla\left(U - \frac{\hbar^2}{4m}\frac{\Psi^*\Delta\Psi + \Psi\Delta\Psi^*}{\rho}\right) \qquad (Ap.\ 1.5)$$

The expression with wave functions in the right part of this equation should be substituted by expressions depending on the probability density. For this purpose we calculate:

$$\Delta\rho = \Psi\Delta\Psi^* + \Psi^*\Delta\Psi + 2\nabla\Psi\nabla\Psi^*$$

$$(\nabla\rho)^2 = (\Psi\nabla\Psi^* + \Psi^*\nabla\Psi)^2 = (\Psi\nabla\Psi^* - \Psi^*\nabla\Psi)^2 + 4\varrho\nabla\Psi\nabla\Psi^* = -\frac{4J^2}{\hbar^2} + 4\varrho\nabla\Psi\nabla\Psi^*$$

Excluding with help of above formulae the expression with wave functions in (Ap. 1.5), we finally get the following equation:

$$\frac{\partial \left(\frac{J}{\rho}\right)}{\partial t} = -\nabla\left(U - \frac{\hbar^2}{4m}\frac{\Delta\rho}{\rho} + \frac{J^2}{2m\rho^2} + \frac{\hbar^2(\nabla\rho)^2}{8m\rho^2}\right),$$ (Ap. 1.6)

which coincides with Equation (3.6). If there is a macroscopic momentum **P** at infinite motion, we substitute $\frac{J}{\rho} = P$ in Equation (Ap. 6) and obtain Expression (3.5).

$$\frac{\partial \mathbf{P}}{\partial t} = -\nabla\left(\frac{P^2}{2m} + U + \frac{\hbar^2(\nabla\rho)^2}{8m\rho^2} - \frac{\hbar^2\Delta\rho}{4m\rho}\right)$$

## Appendix 2
## Quantum Particles Motion in Stationary External Fields

Let us consider a one-dimensional case of quantum particle motion in a stationary external field *U(x)*. The quantum particle is in a needle state when transversal components of quantum energy are equal to zero. The problem is to understand to what extend the solution for a free particle can be used in weak-gradient fields. It is necessary to solve the following system of equations:

$$m\frac{\partial \varrho}{\partial t} + \frac{\partial \varrho P}{\partial x} = 0$$ (Ap. 2.1)

Here $P = P_x(x)$.

$$E = \frac{P^2}{2m} + U + \varepsilon(\rho) = const$$ (Ap. 2.2)

$$\varepsilon(\rho) = \frac{(\delta P)^2}{2m} = \frac{\hbar^2}{8m\varrho^2}\left(\frac{\partial \varrho}{\partial x}\right)^2 - \frac{\hbar^2}{4m\varrho}\frac{\partial^2 \varrho}{\partial x^2}$$ (Ap. 2.3)

We re-arrange Equation (Ap. 2.1) in the form:

$$\frac{m}{P}\frac{\partial \rho P}{\partial t} + \frac{\partial \rho P}{\partial x} = 0$$

We calculate this equation to within $\frac{dP}{dx} \to 0$. Then

$$\varrho P = J\left(\frac{\int \frac{m}{P}dx - t}{T}\right)$$ (Ap. 2.4)

where *T* is the specific problem time. Let us designate

$$\frac{\int \frac{m}{P}dx - t}{T} = \Phi$$

We substitute solution (П2.4) into (П2.3) and obtain:

$$\frac{(\delta P)^2}{2m} \cong \frac{m\hbar^2}{8T^2P^2}\left[\frac{1}{J^2}\left(\frac{dJ}{d\Phi}\right)^2 - \frac{2}{J}\frac{d^2J}{d\Phi^2}\right]$$ (Ap. 2.5)

The satisfying solution has the form:

$$J = J_0 \cos^2\left(\frac{\int \frac{m}{P} dx - t}{T}\right) \quad \text{(Ap. 2.6)}$$

Substituting (Ap. 2.6) into (Ap. 2.5) we get:

$$T = \frac{m\hbar}{P\delta P} = const \quad \text{(Ap. 2.7)}$$

Taking into account that

$$\frac{(\delta P)^2}{2m} = \frac{(\rho\pi\hbar)^2}{8m}$$

We obtain from (Ap. 2.7):

$$P(x)\varrho(x) = \frac{2m}{\pi T} = const$$

In the approximation considered the probability flow density in the coordinate space remains constant.

Thus, finally we find:

$$\varrho = \frac{J_0}{P(x)} \cos^2\left(\frac{\int \frac{m}{P} dx - t}{T}\right) \quad \text{(Ap. 2.8)}$$

$$E = \frac{P(x)^2}{2m} + U(x) + \frac{m\hbar^2}{2T^2 P^2} \quad \text{(Ap. 2.9)}$$

The solutions obtained are similar to the quasi-classical approximation, which is used in traditional quantum mechanics, in the method of derivation. This approximation is true under the following condition:

$$\left|\frac{JT \frac{dP}{dx}}{m \frac{dJ}{d\omega}}\right| \ll 1 \quad \text{(Ap. 2.10)}$$

## Appendix 3
### Solving Quantum Hydrodynamic Equations for Free Particle

Let us consider motion of a free particle with mass $m$ and given momentum $P$. The system of Equations (4.2) and (4.3) takes on form:

$$m\frac{\partial \rho}{\partial t} + P\nabla\rho = 0 \quad \text{(Ap. 3.1)}$$

$$\delta\varepsilon = \frac{(\delta P)^2}{2m} = \frac{\hbar^2}{8m\rho^2}(\nabla\rho)^2 - \frac{\hbar^2}{4m\rho}\Delta\rho = const \tag{Ap. 3.2}$$

We will look for a general solution to equation (Ap. 3.1) in the following form:

$$\rho(t,r) = \varrho(\varphi), \quad \text{где} \quad \varphi = \delta P(r - Pt/m)/\hbar \tag{Ap. 3.3}$$

Then

$$\frac{\partial\rho}{\partial t} = \frac{d\rho}{d\varphi}\frac{\partial\varphi}{\partial t} = -\frac{d\rho}{d\varphi}\delta P \cdot \frac{P}{m\hbar}$$

$$\nabla\rho = \frac{d\rho}{d\varphi}\nabla\varphi = \frac{d\rho}{d\varphi}\frac{\delta P}{\hbar}$$

Substituting $\frac{\partial\rho}{\partial t}$ and $\nabla\rho$ into Expression (Ap. 3.1) we make sure that (Ap. 3.3) is a solution to this equation. Then we calculate

$$\Delta\rho = \frac{d^2\varrho}{d\varphi^2}\frac{(\delta P)^2}{\hbar^2}$$

Substituting $\nabla\rho$ and $\Delta\rho$ into expression (П3.2) we find

$$4 = \frac{1}{\rho^2}\left(\frac{d\rho}{d\varphi}\right)^2 - \frac{2}{\rho}\frac{d^2\varrho}{d\varphi^2} \tag{Ap. 3.4}$$

To solve this equation we reduce the order of derivatives. We designate:

$$\frac{1}{\rho}\frac{d\rho}{d\varphi} = u \tag{Ap. 3.5}$$

Then

$$\frac{du}{d\varphi} = \frac{1}{\rho}\frac{d^2\varrho}{d\varphi^2} - \frac{1}{\rho^2}\left(\frac{d\rho}{d\varphi}\right)^2$$

Substituting these expressions into (П3.4) we obtain:

$$2\frac{du}{d\varphi} + u^2 + 4 = 0$$

We calculate this equation using the method of separation of variables and get:

$$arctg\frac{u}{2} = -\varphi$$

Substituting the solution into (Ap. 3.5) and integrating again we finally obtain:

$$\rho(t,r) = \rho_0 \cos^2\varphi = \rho_0 \cos^2(\delta P(r - \frac{Pt}{m})/\hbar)$$

The solution obtained matches Formula (4.6).

# Appendix 4
# Charged Particle Motion in Electromagnetic Field

Let us consider the motion of a particle with charge *e* and mass *m* in an arbitrary electromagnetic field in quasi-hydrodynamic representation [1]. We suppose that the electric field intensity is $\mathcal{E}$ and the magnetic field intensity is $\mathcal{H}$. These variables can be expressed in terms of magnetic vector and scalar potentials: *A* and $\phi$:

$$\mathcal{E} = \frac{1}{c}\frac{\partial A}{\partial t} - \nabla\phi \qquad (Ap.\ 4.1)$$

$$\mathcal{H} = rot A \qquad (Ap.\ 4.2)$$

In this case the Hamiltonian operator has the following form [2]:

$$\hat{H} = \frac{1}{2m}(\hat{P} - \frac{e}{c}A)^2 + e\phi + U \qquad (Ap.\ 4.3)$$

Tit has been taken into account here that beside electromagnetic forces there are some other forces described by the force function *U*. Using the Coulomb gauge *divA*=0, the Hamiltonian operator can be rewritten in a different form:

$$\hat{H} = \frac{1}{2m}\hat{P}^2 - \frac{e}{mc}(A\hat{P}) + \frac{e^2}{2mc^2}A^2 + e\phi + U \qquad (Ap.\ 4.4)$$

Let us write necessary initial Scroedinger equations for further calculations with operator (Ap. 4.4).

$$i\hbar\frac{\partial\Psi}{\partial t} = \hat{H}\Psi \qquad -i\hbar\frac{\partial\Psi^*}{\partial t} = \hat{H}\Psi^* \qquad (Ap.\ 4.5)$$

Multiplying the first Schroedinger equation by $\Psi^*$ and the second by $\Psi$ and subtracting the second equation from the first one, we obtain

$$i\hbar\frac{\partial\varrho}{\partial t} = \Psi^*\hat{H}\Psi - \Psi\hat{H}\Psi^* = -\frac{\hbar^2}{2m}(\Psi^*\Delta\Psi - \Psi\Delta\Psi^*) + \frac{ie\hbar}{mc}(\Psi^*A\nabla\Psi + \Psi A\nabla\Psi^*)$$

We designate $\varrho = \Psi\Psi^*$. Let us introduce the probability density flow *J/m*

$$J/m = \frac{i\hbar}{2m}(\Psi\nabla\Psi^* - \Psi^*\nabla\Psi) \qquad (Ap.\ 4.6)$$

Let us denote:

$$R = \frac{e}{c}div A\varrho \qquad (Ap.\ 4.7)$$

Then we obtain:

$$m\frac{\partial\rho}{\partial t} + div\ \mathbf{J} - R = 0$$

Let us introduce a new expression for the momentum flow density *J*

$$J^* = J - \frac{e}{c}\rho A$$

We will obtain the probability density conservation law in the following form:

$$m \frac{\partial \rho}{\partial t} + \text{div } \mathbf{J}^* = 0$$

(Ap. 4.8)

If $\mathbf{J} = \varrho \mathbf{P}$, then $\mathbf{J}^* = \rho\left(\mathbf{P} - \frac{e}{c}\mathbf{A}\right) = \varrho \mathbf{P}^*$, where $\mathbf{P}^*$ is a standard generalized momentum of a particle in the electromagnetic field.

Further, the derivation will be calculated:

$$\frac{\partial(\frac{J}{\rho})}{\partial t} = \frac{i\hbar}{2}\nabla\frac{\partial}{\partial t}(\ln\Psi^* - \ln\Psi) = \frac{i\hbar}{2}\nabla\left(\frac{1}{\Psi^*}\frac{\partial\Psi^*}{\partial t} - \frac{1}{\Psi}\frac{\partial\Psi}{\partial t}\right) = -\frac{1}{2}\nabla\left[\frac{\Psi\hat{H}\Psi^* + \Psi^*\hat{H}\Psi}{\varrho}\right]$$

Here Schroedinger equations written above are used. Substituting the explicit form of Hamiltonian operator, we obtain:

$$\frac{\partial(\frac{J}{\rho})}{\partial t} = -\nabla\left(\frac{e^2}{2mc^2}A^2 + e\phi + U - \frac{2e}{mc}\frac{\mathbf{AJ}}{\rho} - \frac{\hbar^2}{4m}\frac{\Psi^*\Delta\Psi + \Psi\Delta\Psi^*}{\rho}\right)$$

(Ap. 4.9)

The term with wave functions in the right part of this equation should be substituted by expressions dependent on the probability density. For this purpose, we calculate:

$$\Delta\rho = \Psi\Delta\Psi^* + \Psi^*\Delta\Psi + 2\nabla\Psi\nabla\Psi^*$$

$$(\nabla\rho)^2 = (\Psi\nabla\Psi^* + \Psi^*\nabla\Psi)^2 = (\Psi\nabla\Psi^* - \Psi^*\nabla\Psi)^2 + 4\varrho\nabla\Psi\nabla\Psi^* = -\frac{4J^2}{\hbar^2} + 4\varrho\nabla\Psi\nabla\Psi^*$$

Excluding the term with wave functions in (Ap. 4.9), we finally obtain:

$$\frac{\partial(\frac{J}{\rho})}{\partial t} = -\nabla\left(\frac{J^2}{2m\rho^2} + \frac{e^2}{2mc^2}A^2 + e\phi + U - \frac{2e}{mc}\frac{\mathbf{AJ}}{\rho} - \frac{\hbar^2}{4m}\frac{\Delta\rho}{\rho} + \frac{\hbar^2(\nabla\rho)^2}{8m\rho^2}\right),$$

(Ap. 4.10)

The new potential could be introduced

$$U^* = \frac{e^2}{2mc^2}A^2 + e\phi + U - \frac{2e}{mc}\frac{\mathbf{AJ}}{\rho}$$

(Ap. 4.11)

Then we get the equation

$$\frac{\partial}{\partial t}(\mathbf{J}/\rho) = -\nabla\left(\frac{J^2}{2m\rho^2} + U^* + \frac{\hbar^2(\nabla\rho)^2}{8m\rho^2} - \frac{\hbar^2\Delta\rho}{4m\rho}\right)$$

(Ap.4.12)

which is in agreement with (Ap. 1.6).

If there is a macroscopic momentum $\mathbf{J}/\rho = \mathbf{P}$, then we have:

$$\frac{\partial}{\partial t}\mathbf{P} = -\nabla\left(\frac{P^2}{2m} + U^* + \frac{\hbar^2(\nabla\rho)^2}{8m\rho^2} - \frac{\hbar^2\Delta\rho}{4m\rho}\right)$$

(Ap.4.13)

which agrees with (Ap. 1.7), but the motion takes place in the effective force field with the potential $U^*$ from (Ap. 4.11). Let us introduce the expression for the particle velocity $\mathbf{v} = \mathbf{P}/m$, then Eq. (Ap. 4.13) can be rewritten as:

$$\frac{\partial v}{\partial t} + (\boldsymbol{v}\nabla)v = -\frac{1}{m}\nabla\left(U^* + \frac{\hbar^2(\nabla\rho)^2}{8m\rho^2} - \frac{\hbar^2\Delta\rho}{4m\rho}\right) \qquad \text{(Ap. 4.14)}$$

This is a quantum equation for a laminar flow of compressible perfect fluid [1].

Reference.

1. B.V. Alexeev, A.I. Abakumov. On Approach to Solving Schrodinger Equation. Doklady Akademii Nauk. V. **262**, **P.** 1100. 1982
2. D.I. Blokhintsev. Fundamentals of Quantum Mechanics. M.: Nauka 1976. 664p.